\documentclass[aps,twocolumn,showpacs,amssymb,pra]{revtex4-1}
\usepackage{amsmath}
\usepackage{hyperref}
\usepackage{graphicx}
\usepackage{bbm}
\usepackage{color}

\usepackage{bm}

\definecolor{darkgreen}{rgb}{0,0.4,0}

\newcommand{\itheta}{i\theta}

\allowdisplaybreaks

\begin{document}
\title{Strongly repulsive anyons in one dimension}

\author{Florian Lange}
\author{Satoshi Ejima}
\author{Holger Fehske}
\affiliation{Institut f{\"u}r Physik,
             Ernst-Moritz--Arndt Universit{\"a}t Greifswald,
             D-17489 Greifswald,
             Germany}

\begin{abstract}
To explore the static properties of the one-dimensional anyon-Hubbard
model for a mean density of one particle per site, we apply perturbation theory with respect to the ratio between kinetic energy and interaction energy  
in the Mott insulating phase. 
The strong-coupling  results for the ground-state energy, the single-particle
excitation energies, and the momentum distribution functions
up to 6th order in hopping are benchmarked against the numerically 
exact (infinite) density-matrix renormalization group technique.
Since these analytic expressions are valid for any fractional phase
$\theta$ of anyons, they will be of great value for a sufficiently reliable analysis of 
future experiments, avoiding extensive and costly numerical 
simulations.
\end{abstract}

\maketitle

\section{Introduction}

Particles are usually classified as either bosons or fermions, depending on whether the wave function is symmetric or antisymmetric with respect to the exchange of two identical particles. 
Some systems, however, may realize quasiparticles with fractional
statistics, called anyons, that acquire a complex phase factor
$e^{\itheta}$ with $0<\theta<\pi$ under exchange~\cite{LM77,Wi82}. 
Most notably, anyons have been used in the description of the fractional quantum Hall effect~\cite{TSG82,Laughlin83}. 
While anyons are usually restricted to two-dimensional systems, fractional statistics can in principle be defined in arbitrary dimensions~\cite{Haldane91}.

One dimensional (1D) anyon models can be expressed in terms of bosonic operators by using a generalized Jordan--Wigner transformation. 
There are several proposals to utilize this equivalence to implement an anyon-Hubbard model (AHM) by loading ultracold atoms in  optical lattices. The fractional exchange statistics is thereby translated into an occupation-dependent hopping phase that experimentally may be implemented by assisted Raman tunneling~\cite{KLMR11,GS15} or lattice-shaking-assisted tunneling against potential offsets~\cite{SSE16}. One of the advantages of any optical lattice setup is the high degree of control of system parameters including the statistical angle $\theta$. As yet, however, an experimental realization of anyons in optical lattices has not been archived.

Since the  introduction of the AHM \cite{KLMR11}, several theoretical and numerical studies have been carried out, \textit{inter alia},  exploring the effect of fractional statistics on momentum distributions~\cite{TEP15} 
and the position of the quantum phase transition between the Mott insulator (MI) and superfluid (SF)~\cite{KLMR11,AFS16} as well as revealing additional phases such as an exotic two-component partially-paired phase~\cite{GS15}. 
It has been shown that the various superfluid phases of the AHM can be qualitatively understood within a generalized Gutzwiller mean-field ansatz~\cite{TEP15,ZGFSZ16}. 
Here, we instead focus on the MI phase, using strong-coupling perturbation theory as it has been applied to the Bose-Hubbard model (BHM)~\cite{EFGzMKAvdL12,DZ06,EM99b}. 
In addition to the perturbative analysis, we study the model numerically with the density-matrix renormalization group (DMRG)~\cite{White92,Mc08,Sch11} and a variational matrix-product state (MPS) ansatz for dispersion relations~\cite{HPWCOVV12,HOV13}.

The outline of this paper is as follows: In Sec.~\ref{sec:model}
we introduce the 1D AHM and apply the anyon-boson mapping by the
fractional version of the Jordan--Wigner transformation to the AHM
in order to rewrite the Hamiltonian with the bosonic operators.
In Sec.~\ref{sec:SCE} we describe the strong-coupling analysis
for the ground-state energy, the momentum-dependent single-hole
and single-particle excitation energies, and the momentum 
distribution functions. To evaluate the validity of the proposed 
strong-coupling approach we perform  an extensive comparison 
with unbiased data obtained by the MPS-based (infinite) DMRG (iDMRG)  technique. 
Finally, Sec.~\ref{sec:summary} summarizes our results and gives a brief outlook.

\section{Anyonic Hubbard model}
\label{sec:model}
On a linear chain of $L$ sites with periodic boundary conditions 
(PBCs), the Hamiltonian of the 1D AHM is defined as
$\hat{H}_{\rm AHM}^{(a)}\equiv t\hat{T}_a+U\hat{D}$,
with
\begin{eqnarray}
 \hat{T}_a&=&-\sum_{j=1}^{L}
  \left(
   \hat{a}_j^{\dagger}\hat{a}_{j+1}^{\phantom{\dagger}}
   +\hat{a}_{j+1}^{\dagger}\hat{a}_{j}^{\phantom{\dagger}}
  \right) 
 \label{a-hop-term}
 \end{eqnarray}
and
\begin{eqnarray}
\hat{D}&=&
 \frac{1}{2}\sum_{j=1}^{L}\hat{n}_j\left(\hat{n}_j-1\right)\,,
\label{a-d-term}
 \end{eqnarray}
describing the nearest-neighbor anyon transfer ($\propto t$)
and the on-site anyon repulsion ($\propto U$), respectively.
Here, $\hat{a}_j^{\dagger}$, $\hat{a}_j^{\phantom{\dagger}}$,
and $\hat{n}_j=\hat{a}_j^{\dagger}\hat{a}_j^{\phantom{\dagger}}$ are the anyon creation, annihilation, 
and particle number operators on site $j$, respectively, which fulfill  the generalized commutation relations~\cite{KLMR11}
\begin{eqnarray}
 \hat{a}_{j}^{\phantom{\dagger}}\hat{a}_{\ell}^{\dagger}
  -e^{-i\theta\mathrm{sgn}(j-\ell)}
  \hat{a}_{\ell}^{\dagger}\hat{a}_{j}^{\phantom{\dagger}}
  &=&\delta_{j\ell}\, ,
  \\
 \hat{a}_{j}\hat{a}_{\ell}
  -e^{i\theta\mathrm{sgn}(j-\ell)}a_{\ell}a_{j}
  &=& 0\, .
 \label{CCR}
\end{eqnarray}
Since $\mathrm{sgn}(0)=0$, regular bosonic commutation relations apply for particles on the same site. Anyons with the fractional angle
$\theta=\pi$ represent the so-called ``pseudofermions,'' namely,
they behave as ordinary fermions off-site, while being
bosons on-site.

Carrying out a fractional Jordan--Wigner transformation~\cite{KLMR11},
\begin{equation}
 \hat{a}_j=
  \hat{b}_j e^{i\theta\sum_{\ell=1}^{j-1}\hat{n}_\ell}\,,
  \label{ab-mapping}
\end{equation}
$\hat{T}_a$ of Eq.~\eqref{a-hop-term} can be rewritten
with boson creation ($\hat{b}_j^\dagger$) and annihilation
($\hat{b}_j^{\phantom{\dagger}}$) operators as
\begin{eqnarray}
 \hat{T}_b=-\sum_j
  \left(
   \hat{b}_j^{\dagger}\hat{b}_{j+1}^{\phantom{\dagger}}
   e^{i\theta\hat{n}_j}
   +e^{-i\theta\hat{n}_j}
   \hat{b}_{j+1}^{\dagger}\hat{b}_{j}^{\phantom{\dagger}}
  \right)\, .
  \label{Tb}
\end{eqnarray}
To be more precise, when an anyon hops to the left from
site $j+1$ to site $j$, an occupation dependent phase
$e^{\itheta\hat{n}_j}$ is picked up in the bosonic operator
description. Note that
$\hat{n}_j=\hat{a}_j^\dagger\hat{a}_j^{\phantom{\dagger}}
=\hat{b}_j^\dagger\hat{b}_j^{\phantom{\dagger}}$,
so that the on-site repulsion term $\hat{D}$ is form-invariant under the anyon-boson mapping~\eqref{ab-mapping}. 
In order to study the model deep in the Mott-insulating regime, we apply an $x=t/U$ strong-coupling expansion to $\hat{H}_{\rm AHM}^{(b)}=t\hat{T}_b+U\hat{D}$. Throughout this work, we restrict ourselves to unit filling. 
\section{Strong-coupling expansions}
\label{sec:SCE}
\subsection{Ground state}
At integer filling $\rho=N/L$, the AHM has a unique ground state,  
\begin{equation}
 |\phi_0\rangle= \frac{1}{(\rho !)^{L/2}}
 \prod_{i} \left(\hat{b}_i^{\dagger}\right)^{\rho} |{\rm vac}\rangle\;,
 \label{phi0}
\end{equation}
in the limit $x\to0$. The state $|\phi_0\rangle$ can be used as a starting point for a perturbative calculation of the ground state in the MI phase.

Executing the unitary Harris-Lange transformation~\cite{HL67},
the strong-coupling Hamiltonian of the AHM is derived 
in a similar way as for the (BHM)~\cite{EFGzMKAvdL12}:
\begin{eqnarray}
\hat{h}&=&e^{\hat{S}} \hat{H} e^{-\hat{S}} 
= U \hat{D} +t \sum_{r=0}^{\infty}x^r \hat{h}_r \; ,\\
\hat{S}&=&-\hat{S}^{\dagger}=
\sum_{r=1}^{\infty}x^r \hat{S}_r \; .
\end{eqnarray}
Practically, we keep a finite order in the expansion of $\hat{S}$.
Retaining $\hat{S}_r$ for $1\leq r\leq n$ denotes the ``$n$th-order expansion.'' 
The operators $\hat{S}_n$ are defined by requiring that in the $n$th order for $S$, the transformed Hamiltonian conserves the number of double occupancies to 
$(n-1)$th order, that is, $[\hat{h}_r, \hat{D}]=0$ for $1\leq r\leq n-1$. 
Higher-order terms in the expansion of $\hat{h}$ are neglected, so $|\phi_0\rangle$ is an eigenstate of the strong-coupling Hamiltonian.

Following this recipe, the leading-order terms 
for $\hat{S}_r$ and $\hat{h}_r$ are obtained as
\begin{eqnarray}
\hat{S}_1&=&\sum_{D_1,D_2} 
\frac{\hat{P}_{D_1} \hat{T} \hat{P}_{D_2}}{D_1-D_2}\; ,
\\[6pt]
\hat{S}_2&=&\sum_{D_1,D_2} 
\frac{-\hat{P}_{D_1} \hat{T} \hat{P}_{D_1} \hat{T} \hat{P}_{D_2}
+\hat{P}_{D_1} \hat{T} \hat{P}_{D_2} \hat{T} \hat{P}_{D_2}}{(D_1-D_2)^2}
\\[3pt]
&& +\!\! \sum_{D_1,D_2,D_3}\!\!
\frac{\hat{P}_{D_1} \hat{T} \hat{P}_{D_3} \hat{T} \hat{P}_{D_2}}{2(D_1-D_2)}
\frac{[D_1-D_3+D_2-D_3]}{(D_1-D_3)(D_2-D_3)}
\nonumber 
\; ,
\\[6pt]
\hat{h}_0&=& \sum_{D}\hat{P}_D \hat{T} \hat{P}_{D} \; ,
\label{hath0}
\\[6pt]
\hat{h}_1&=& \sum_{D_1,D_2}
\frac{\hat{P}_{D_1} \hat{T} \hat{P}_{D_2} \hat{T} \hat{P}_{D_1}}{D_1-D_2}\; ,
\end{eqnarray}
where $\hat{P}_D$ is the projection operator onto the subspace
of eigenstate with $D$ interactions, $\hat{D}=\sum_{D=0}^{\infty} D\hat{P}_D$.
In the above sums it is implicitly suggested that all indices 
$D_i\geq 0$ are different from each other. Higher orders are generated recursively as described in Ref.~\cite{vanDongen94}, where the necessary bookkeeping can be done by a computer algebra program. The resulting expansion differs from the one for the BHM only by the hopping operator $\hat{T}$. 

Within the strong-coupling expansion the ground state $|\psi_0\rangle$ and ground-state energy $E_0$ of the original Hamiltonian are
\begin{eqnarray}
 |\psi_0\rangle = e^{\hat{S}}|\phi_0\rangle\,, \quad
\hat{h}|\phi_0\rangle = E_0 |\phi_0\rangle \,,
\end{eqnarray}
where $|\phi_0\rangle$ is the ground state of $\hat{h}_{-1}=\hat{D}$,
see Eq.~\eqref{phi0}. Since the Harris--Lange transformation is unitary
the operators and ground-state expectation values are translated 
with $|\psi_0\rangle \mapsto  |\phi_0\rangle$,  $\hat{H} \mapsto  \hat{h}$, and $\hat{A} \mapsto  \widetilde{A}=e^{\hat{S}} \hat{A} e^{-\hat{S}} $. 
 
Calculating the various observables in the strong-coupling expansion then amounts to evaluating chains of hopping operators in the unperturbed ground state $|\phi_0\rangle$ which are weighted depending on how they change the number of double occupancies at each step. In doing so, one only has to sum over connected hopping processes that can be {evaluated using finite clusters. 
The difference between the AHM and the BHM enters the strong-coupling expansion through the phase factors picked up by the hopping processes or an explicit $\theta$-dependence of the observables.

The ground-state energy is simply given by 
\begin{equation}
 E_0=\langle\phi_0|\hat{h}|\phi_0\rangle \;.
\end{equation}
Up to 6th order in $x$, we obtain for the rescaled ground-state energy per site
\begin{eqnarray}
 \frac{E_0^{[6]}}{4UL}&=&
  -x^2 +\left(6-5\cos(\theta)\right)x^4  
  \nonumber\\
 &&
  +\frac{1}{9}\left(-872 + 1168\cos(\theta) 
    - 228\cos(2\theta)\right)x^6
  \nonumber\\
 && +O(x^8)
  \label{e0-sc}
\end{eqnarray}
in agreement with Refs.~\cite{DZ06,EFGzMKAvdL12} in the BHM limit
$\theta\to 0$.

\begin{figure}[tb]
\begin{center}
 \includegraphics[width=0.9\linewidth,clip]{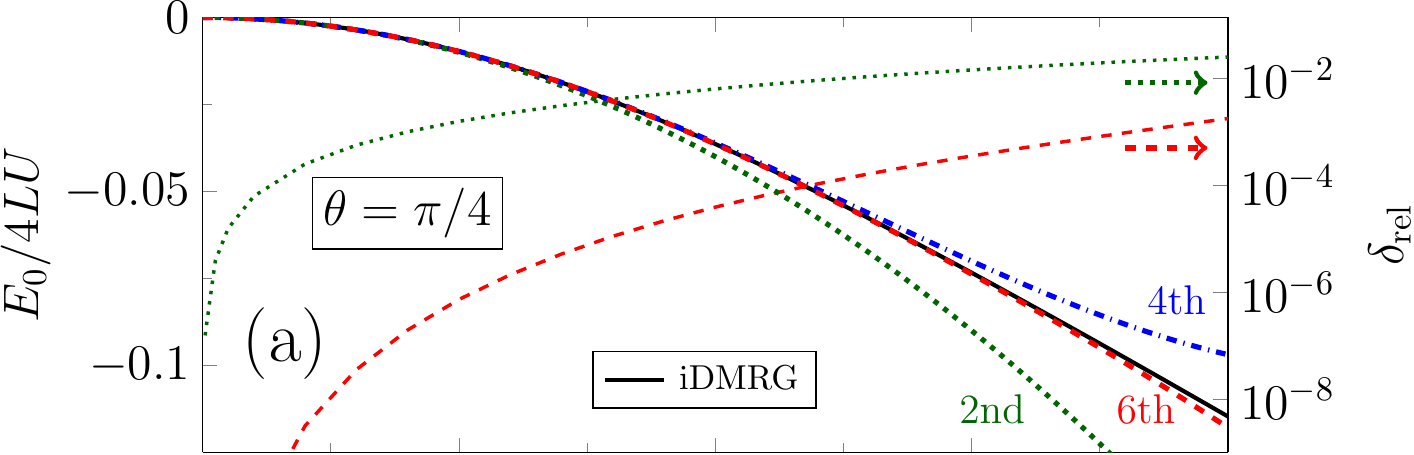}
 \includegraphics[width=0.9\linewidth,clip]{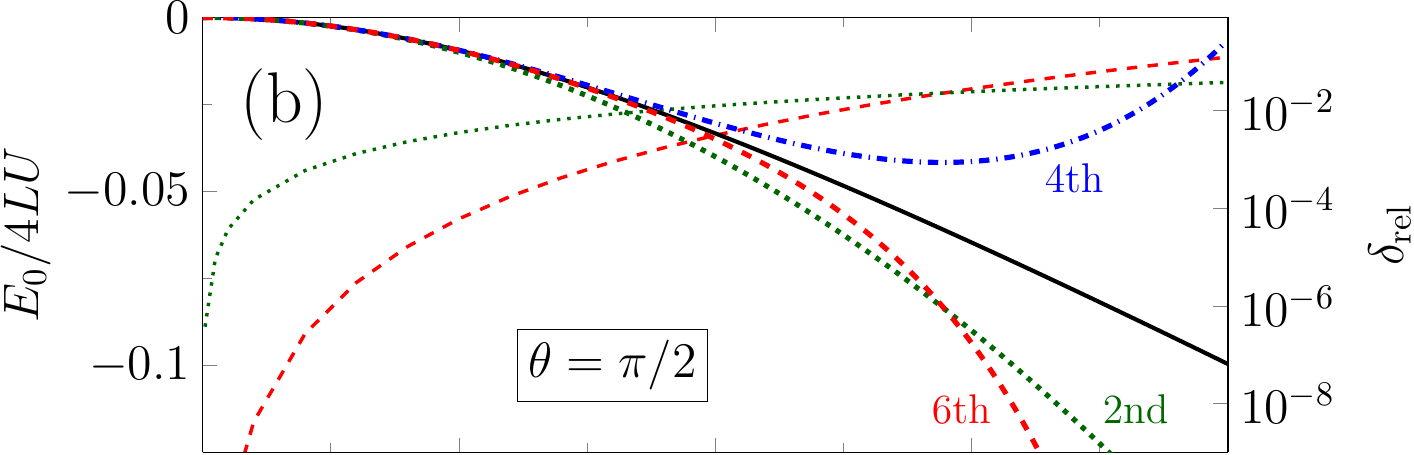}
 \includegraphics[width=0.9\linewidth,clip]{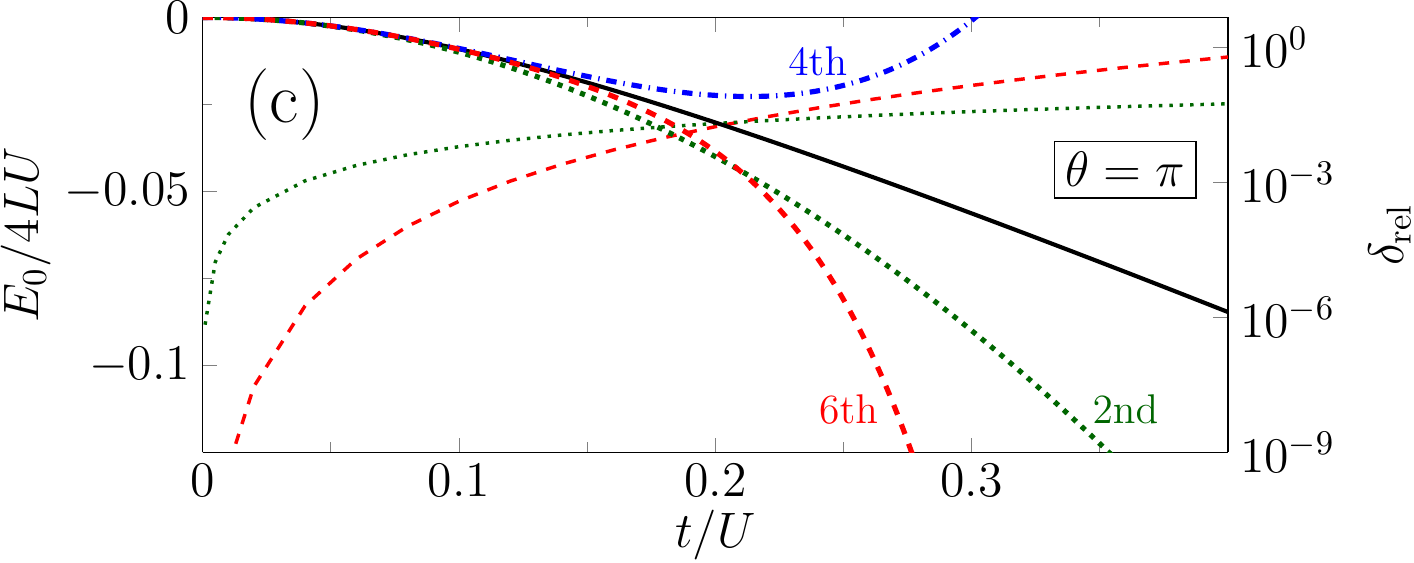}
\end{center}
 \caption{(Color online) Ground-state energy $E_0/4LU$ as a function 
 of interaction strength $t/U$. 
 $n$th-order strong-coupling results $E_0^{[n]}$ of Eq.~\eqref{e0-sc}
 are compared with the quasi-exact iDMRG data $E_0^{\rm ex}$ for $\chi=100$
 (solid lines). 
The relative errors $\delta_{\rm rel}=|(E_0^{\rm ex}-E_0^{[n]})/E_0^{\rm ex}|$ are given
 in semi-logarithmic representation corresponding to the right $y$-axis.
 \label{fig:e0}}
\end{figure}

Figure~\ref{fig:e0} compares the strong-coupling perturbation theory
with iDMRG results for various $\theta$.
Similar to the case in the BHM~\cite{EFGzMKAvdL12}, for small $\theta$ 
[e.g., Fig.~\ref{fig:e0}(a) for $\theta=\pi/4$], the strong-coupling 
series expansion is
in reasonable accordance
with the numerically exact result,
as indicated clearly by the relative errors in Fig.~\ref{fig:e0}.
For 6th order in $x$, the deviation starts in the intermediate-coupling regime 
at $t/U\simeq 0.35$. 
As expected, the quality of the perturbation analysis improves 
as higher-order corrections are taken into account. This is valid
for all $\theta$ as demonstrated in Fig.~\ref{fig:e0}.

Figure~\ref{fig:e0} also shows that the range of validity of the 
strong-coupling theory becomes worse with increasing $\theta$.
The deviation starts already at $t/U\simeq 0.12$ in the case of 
$\theta=\pi$, see panel~(c).

\subsection{Excitation energies}
\label{sec:ep-eh}
Similar to the ground state $|\phi_0\rangle$, Eq.~\eqref{phi0},
the energy levels of a single-hole
excitation, $E_{\rm h}(k)$, and of a single-particle excitation,
$E_{\rm p}(k)$, can 
be extracted from the strong-coupling
expansion to high order in $x$, since the perturbation analysis
for these energy levels also starts from nondegenerate states, 
i.e., in the case of $\rho=1$, 
\begin{eqnarray}
 |\phi_{\rm h}(k)\rangle&=& 
\sqrt{\frac{1}{L}}
\sum_{\ell=1}^L e^{-{i}k\ell} \hat{b}_\ell |\phi_0\rangle\; ,
\label{phi-hole}\\
|\phi_{\rm p}(k)\rangle&=& \sqrt{\frac{1}{L}}\sqrt{\frac{1}{2}}
\sum_{\ell=1}^L e^{{i}k\ell} \hat{b}_\ell^{\dagger} |\phi_0\rangle\; .
\label{phi-particle}
\end{eqnarray}
Therefore, the single-hole and single-particle excitation energies
can be obtained from
\begin{eqnarray}
E_{\rm h}(k) &=& 
 \langle \phi_{\rm h}(k) |\hat{h}|\phi_{\rm h}(k)\rangle 
 - E_0\; ,
 \\
 E_{\rm p}(k) &=& \langle \phi_{\rm p}(k) 
|\hat{h}|\phi_{\rm p}(k)\rangle -E_0\; .
\end{eqnarray}
Carrying out the above perturbation analysis up to and 
including 6th order in $x$, 
we obtain 
\begin{widetext}
\begin{eqnarray}
\frac{E_{\rm h}^{[6]}(k)}{t} &=&  
- 2 \cos{ ( k )} +x\{ 8 - 4 \cos{(2 k + \theta )} \}
\nonumber\\
 && + x^{2} \{
  8 \cos{\left (k \right )} 
  + 4 \cos{\left (k + \theta \right)} 
  - 4 \cos{\left (3 k + \theta \right )}
  - 8 \cos{\left (3 k + 2 \theta \right )} 
  \}
\nonumber\\
 && + x^{3} \{
  - 56 +56 \cos{\left (\theta \right )} 
  + 64 \cos{\left (2 k + \theta \right )} 
  - 4 \cos{\left (4 k + \theta \right )} 
  - 24 \cos{\left (4 k + 2 \theta \right )} 
  - 16 \cos{\left (4 k + 3 \theta \right )} 
  \}
\nonumber\\
 && + x^{4} \biggl\{
  - \frac{256}{3} \cos{\left (k \right )} 
  + \frac{88}{3} \cos{\left (k - \theta \right )} 
  - \frac{104}{3} \cos{\left (k + \theta \right )} 
  + 16 \cos{\left (k + 2 \theta \right )} 
  - 8 \cos{\left (3 k \right )} 
  + 100 \cos{\left (3 k + \theta \right )} 
  \nonumber\\
 && \phantom{+ x^{4} \biggl\{}
  + 216 \cos{\left (3 k + 2 \theta \right )} 
  - 32 \cos{\left (3 k + 3 \theta \right )} 
  - 4 \cos{\left (5 k + \theta \right )} 
  - 48 \cos{\left (5 k + 2 \theta \right )} 
  - 96 \cos{\left (5 k + 3 \theta \right )} 
  \nonumber\\
 && \phantom{+ x^{4} \biggl\{}
  - 32 \cos{\left (5 k + 4 \theta \right )}
  \biggr\}
\nonumber\\
 && + x^{5} \biggl\{
    \frac{2896}{3}
  - \frac{4480}{3} \cos{\left (\theta \right )} 
  + \frac{1072}{3} \cos{\left (2 \theta \right )} 
  + \frac{784}{3} \cos{\left (2 k \right )} 
  + 28 \cos{\left (2 k - \theta \right )} 
  - \frac{3392}{3} \cos{\left (2 k + \theta \right )} 
  \nonumber\\
 && \phantom{+ x^{5} \biggl\{}
  + \frac{928}{3} \cos{\left (2 k + 2 \theta \right )} 
  + \frac{160}{3} \cos{\left (2 k + 3 \theta \right )} 
  - 16 \cos{\left (4 k \right )} 
  + \frac{208}{3} \cos{\left (4 k + \theta \right )} 
  + 864 \cos{\left (4 k + 2 \theta \right )} 
  \nonumber\\
 && \phantom{+ x^{5} \biggl\{}
  + \frac{1520}{3} \cos{\left (4 k + 3 \theta \right )} 
  - 128 \cos{\left (4 k + 4 \theta \right )} 
  - 4 \cos{\left (6 k + \theta \right )} 
  - 80 \cos{\left (6 k + 2 \theta \right )} 
  - 320 \cos{\left (6 k + 3 \theta \right )} 
  \nonumber\\
 && \phantom{+ x^{5} \biggl\{}
  - 320 \cos{\left (6 k + 4 \theta \right )} 
  - 64 \cos{\left (6 k + 5 \theta \right )} 
  \biggr\}
\label{Eh}
\end{eqnarray} 
and
\begin{eqnarray}
 \frac{E_{\rm p}^{[6]}(k)}{t} &=&  
  \frac{1}{x}- 4 \cos{\left (k + \theta \right )} 
  + x \{5  - 4 \cos{\left (2 k + \theta \right )}\} 
  \nonumber\\
 && + x^{2} 
  \left\{
   8 \cos{\left (k \right )} 
   + 22 \cos{\left (k + \theta \right )} 
   - 12 \cos{\left (k + 2 \theta \right )} 
   - 4 \cos{\left (3 k + \theta \right )}
   - 8 \cos{\left (3 k + 2 \theta \right )} 
  \right\}
  \nonumber\\
 && + x^{3} \biggl\{
  - \frac{1969 }{20}
  + \frac{364 }{5} \cos{\left (\theta \right )} 
  + 70  \cos{\left (2 k + \theta \right )} 
  - 18  \cos{\left (2 k + 2 \theta \right )} 
  + 12  \cos{\left (2 k + 3 \theta \right )} 
  - 4  \cos{\left (4 k + \theta \right )} 
  \nonumber\\
 && \phantom{+ x^{3} \biggl\{}
  - 24  \cos{\left (4 k + 2 \theta \right )} 
  - 16  \cos{\left (4 k + 3 \theta \right )} 
  \biggr\}
  \nonumber\\
 && + x^{4}\biggl\{
  - \frac{13384}{75} \cos{\left (k \right )} 
  + 84 \cos{\left (k - \theta \right )} 
  + \frac{2189}{150} \cos{\left (k + \theta \right )} 
  + \frac{10331}{75} \cos{\left (k + 2 \theta  \right )} 
  - \frac{294}{5} \cos{\left (k + 3 \theta \right )} 
  \nonumber\\
 && \phantom{+ x^{4} \biggl\{}
  - 8 \cos{\left (3 k \right )}
  + 102 \cos{\left (3 k + \theta \right )} 
  + 204 \cos{\left (3 k + 2 \theta \right )} 
  - 10 \cos{\left (3 k + 3 \theta \right )} 
  - 12 \cos{\left (3 k + 4 \theta \right )} 
  \nonumber\\
 && \phantom{+ x^{4} \biggl\{}
  - 4 \cos{\left (5 k + \theta \right )} 
  - 48 \cos{\left (5 k + 2 \theta \right )} 
  - 96 \cos{\left (5 k + 3 \theta \right )} 
  - 32 \cos{\left (5 k + 4 \theta \right )}
  \biggr\}
  \nonumber\\
 && + x^{5}\biggl\{
    \frac{794483}{600}
    - \frac{101513}{50} \cos{\left (\theta \right )} 
    + \frac{22907}{75} \cos{\left (2 \theta \right )} 
    + \frac{14296}{75} \cos{\left (2 k \right )} 
    + 68 \cos{\left (2 k - \theta \right )} 
    - \frac{668719}{750} \cos{\left (2 k + \theta \right )} 
  \nonumber\\
 && \phantom{+ x^{5} \biggl\{}
    - \frac{14284}{375} \cos{\left (2 k + 2 \theta \right )} 
    + \frac{1896}{125} \cos{\left (2 k + 3 \theta \right )} 
    + \frac{3663}{25} \cos{\left (2 k + 4 \theta \right )} 
    - \frac{294}{5} \cos{\left (2 k + 5 \theta \right )} 
    - 16 \cos{\left (4 k \right )} 
  \nonumber\\
 && \phantom{+ x^{5} \biggl\{}
    + 70 \cos{\left (4 k + \theta \right )} 
    + 858 \cos{\left (4 k + 2 \theta \right )} 
    + 526 \cos{\left (4 k + 3 \theta \right )} 
    - 154 \cos{\left (4 k + 4 \theta \right )} 
    + 12 \cos{\left (4 k + 5 \theta \right )} 
  \nonumber\\
 && \phantom{+ x^{5} \biggl\{}
    - 4 \cos{\left (6 k + \theta \right )} 
    - 80 \cos{\left (6 k + 2 \theta \right )} 
    - 320 \cos{\left (6 k + 3 \theta \right )} 
    - 320 \cos{\left (6 k + 4 \theta \right )} 
    - 64 \cos{\left (6 k + 5 \theta \right )} 
 \biggr\}\, .
\label{Ep}
\end{eqnarray}
\end{widetext}
In the BHM limit ($\theta=0$), we obtain 
Eqs.~\eqref{Eh-BHM} and \eqref{Ep-BHM}, 
which agree with Eqs.~(24) and (25) of Ref.~\cite{EFGzMKAvdL12}, 
respectively, when correcting some misprints; see App.~\ref{sec:BH-limit}. 

\begin{figure*}[t!]
\begin{center}
\includegraphics[width=0.95\linewidth,clip]
 {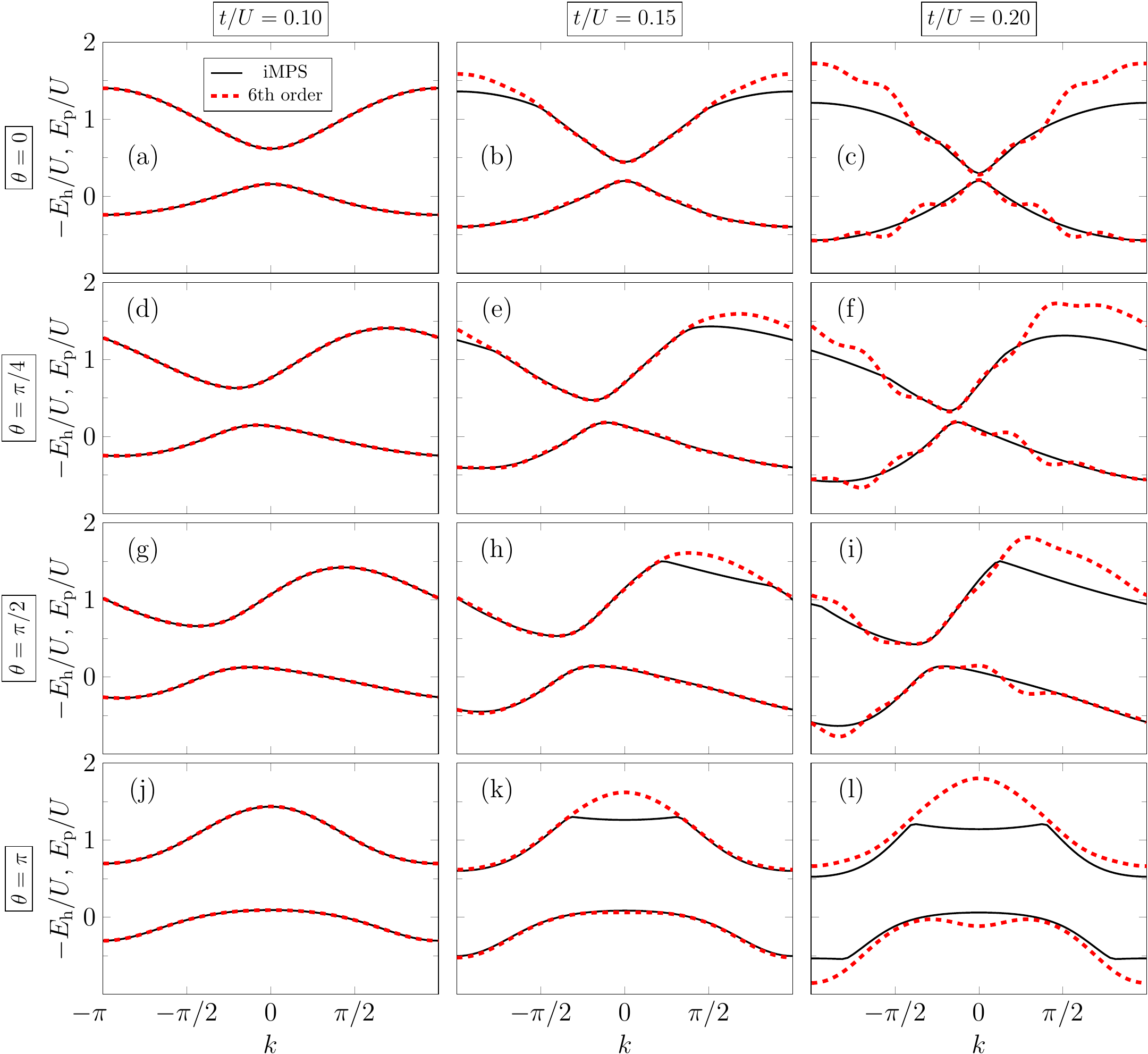}
\end{center}
\caption{(Color online) Sixth-order strong-coupling expansions (dashed lines)
 of the single-hole and single-particle excitation energies, 
 Eqs.~\eqref{Eh} and \eqref{Ep}, compared with numerical data by iMPS 
 with the variational ansatz (solid lines).
}
\label{gap-k}
\end{figure*}
To calculate the dispersion relations of the particle and hole 
excitations numerically, we use the variational MPS ansatz introduced 
in Refs.~\cite{HPWCOVV12,HOV13} that works directly 
in the thermodynamic limit. In the following, we give a rough
description of the method. Starting point is an infinite MPS (iMPS) 
approximation of the ground state
\begin{equation}
 |\psi_0\rangle = v_{\text{L}}^{\dagger} 
  \left( \prod_{j\in \mathbb{Z}} \sum_{s_j} A^{s_j} \right) 
  v_{\text{R}} |\bm{s}\rangle ,
\end{equation}
where $|\bm{s}\rangle = |...,s_j,s_{j+1},...\rangle$, 
the indices $s_n$ label the states of the local Hilbert spaces, 
$A^{s}$ are site independent $\chi \times \chi$ complex matrices, 
and $v_{\text{L}}$ and $v_{\text{R}}$ are $\chi$-dimensional vectors. The boundary vectors $v_{\text{L}}$ and $v_{\text{R}}$ will not affect the bulk properties and can therefore be ignored. 
It is assumed that the transfer matrix $\sum_s A^s \otimes \bar{A}^{s}$ 
has one eigenvalue $1$ and that its other eigenvalues are smaller 
in magnitude.
To calculate $|\psi_0\rangle$, we use the iDMRG.
The ansatz for the elementary excitations is a momentum superposition 
of local perturbations which are introduced by replacing the matrices 
$A^s$ at a single site with matrices $B^s$:
\begin{equation}
| \phi_k(B) \rangle = \sum_{j \in \mathbb{Z}}  
 e^{ikj} \sum_{\{s\}} v_{\text{L}}^{\dagger} 
 (... A^{s_{j-1}} B^{s_j} A^{s_{j+1}} ...) v_{\text{R}}|\bm{s}\rangle.
\end{equation} 
This includes all excitations that are induced by one-site operators 
but can also describe, to some degree, those corresponding 
to operators with larger support.  
Increasing the bond dimension $\chi$ of $|\psi_0\rangle$ results, 
in addition to a better approximation for the ground state energy, 
in a more general ansatz for the excitations. 
One can define matrices $\mathsf{N}_k$ and $\mathsf{H}_k$ such that
\begin{align}
 \langle \phi_k(B) | \phi_{k'}(B') \rangle &=
 2\pi \delta(k-k') B^{\dagger}\mathsf{N}_kB', 
 \\
 \langle \phi_k(B) | \hat{H} - E_0 |\phi_{k'}(B') \rangle
 &= 2\pi \delta(k-k') B^{\dagger}\mathsf{H}_kB',
\end{align}
where $E_0$ is the (infinite) ground-state energy 
and the matrices 
$B^s$ have been combined and reshaped into a vector. 
The approximate excitation energies for any momentum $k$ 
can then be obtained by solving the generalized eigenvalue problem 
for the effective Hamiltonian $\mathsf{H}_k$ and 
the normalization matrix $\mathsf{N}_k$. 
As described in detail in Ref.~\cite{HOV13}, $B$ must be appropriately 
parameterized to exclude zero modes which would result 
in $|\phi_k(B)\rangle = 0$ and to impose orthogonality 
to the ground state. 
A linear parametrization fulfilling these requirements 
can be chosen such that the normalization matrix becomes 
the identity and only a regular eigenvalue problem needs to be solved. 
Since the number of particles is a good quantum number, 
we can separately target particle and hole excitations 
to obtain both $E_{\text{p}}$ and $E_{\text{h}}$.


In Fig.~\ref{gap-k} we compare the strong-coupling results
up to 6th order in $x$ with the lowest excitation energy obtained
by the above mentioned iMPS technique. First of all, $E_{\rm h/p}$ are 
clearly symmetric about $q=0$ in the BHM limit $\theta=0$,
although they become asymmetric for $0<\theta<\pi$ reflecting the 
influence of the fractional angle $\theta$. By considering the
strong-coupling expansions up to 1st order only, this asymmetry 
of the excitation energies can be understood well: the minimum
of excitation energy, $\min\{E_{\rm p}(k)\}$ ($\min\{E_{\rm h}(k)\}$),
shifts from $k=0$ to $-\pi<k<0$, consistent with 
the positive sign of $\theta$ in cosine terms up to 
1st order. Quantitatively, the 6th-order expansions agree perfectly 
with iMPS data up to $x\lesssim0.1$. The deviation between both
results starts about $x\sim 0.15$ especially in $E_{\rm p}(k)$. 
The single-particle excitation from the perturbation theory 
is clearly higher than the lowest excitation energy by iMPS,
e.g. for $-\pi<k<-\pi/2$ and $\pi/2<k<\pi$ with $\theta=0$.
Most probably, the lowest excitation by iMPS stems from a 
many-particle excitation such as two particles and one hole that are forced into an artificial bound state by the iMPS ansatz~\cite{HOV13}. 
Moreover, plotting the higher excitation energies, it is obvious that a continuum of excitations starts to 
arise in this regime.
Figure~\ref{gap-k-t015quarter} demonstrates a typical example
for $E_{\rm p}(k)$ with the parameter sets of Fig.~\ref{gap-k}.
Because of the finite bond dimension $\chi$, the continuous part of the spectrum is approximated by a finite number of discrete energy levels. 
With increasing $x$ further, the results of 
strong-coupling expansions start to oscillate, see Fig.~\ref{gap-k}
for $x\gtrsim 0.2$. 

\begin{figure}[tb]
\begin{center}
\includegraphics[width=0.95\linewidth,clip]
 {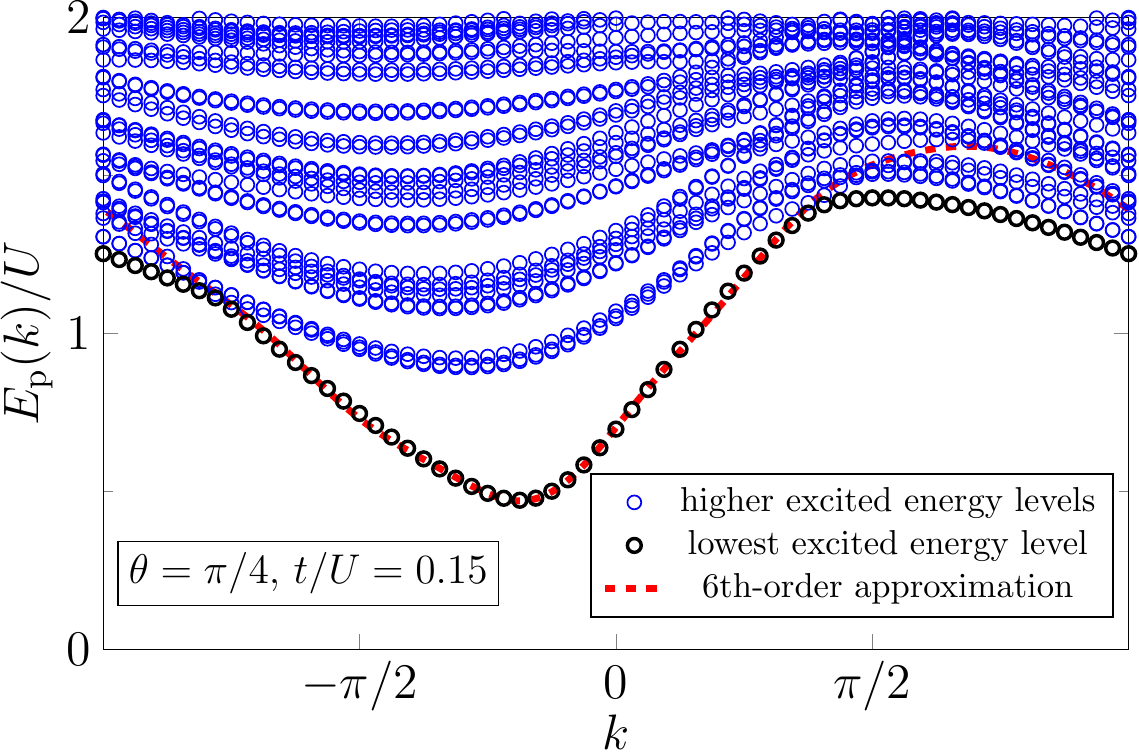}
\end{center}
\caption{(Color online) Excitation energies for one additional particle, the parameters being the same as in panel (e) of Fig.~\ref{gap-k}. The disagreement between perturbation theory and variational iMPS ansatz can be explained by the onset of the multi-particle continuum.    
}
\label{gap-k-t015quarter}
\end{figure}

From the single-hole and single-particle dispersions we can obtain the phase boundaries between MI and SF in the grand-canonical ensemble. 
The chemical potentials $\mu^{\pm}$ at which the phase transitions for fixed $t/U$ take place are determined by the minimum energies for adding a particle or hole to the MI ground state: $\mu^+=\min\{E_{\rm p}(k)\}$ and $\mu^-=-\min\{E_{\rm h}(k)\}$. 
In general, the minima of the strong-coupling expressions~\eqref{Eh} and \eqref{Ep} have to be found numerically. However, in the BHM limit ($\theta=0$) we have  
$\mu^+=E_{\rm p}(0)$ and $\mu^-=-E_{\rm h}(0)$ and thus 
the gap is given by $\Delta=\mu^+-\mu^-=E_{\rm p}(0)+E_{\rm h}(0)$. 
In this way, we can reproduce the single-particle gap in the BHM 
\begin{equation}
\frac{\Delta}{U}=1-6x+5x^{2}+6x^{3}+\frac{287}{20}x^{4}
+\frac{5821}{50}x^{5} -\frac{602243}{1000}x^{6} 
+ \ldots \; ,
\end{equation}
in agreement with Ref.~\cite{EM99b}.

\begin{figure*}[t!]
 \begin{center}
  \includegraphics[width=0.95\linewidth,clip]{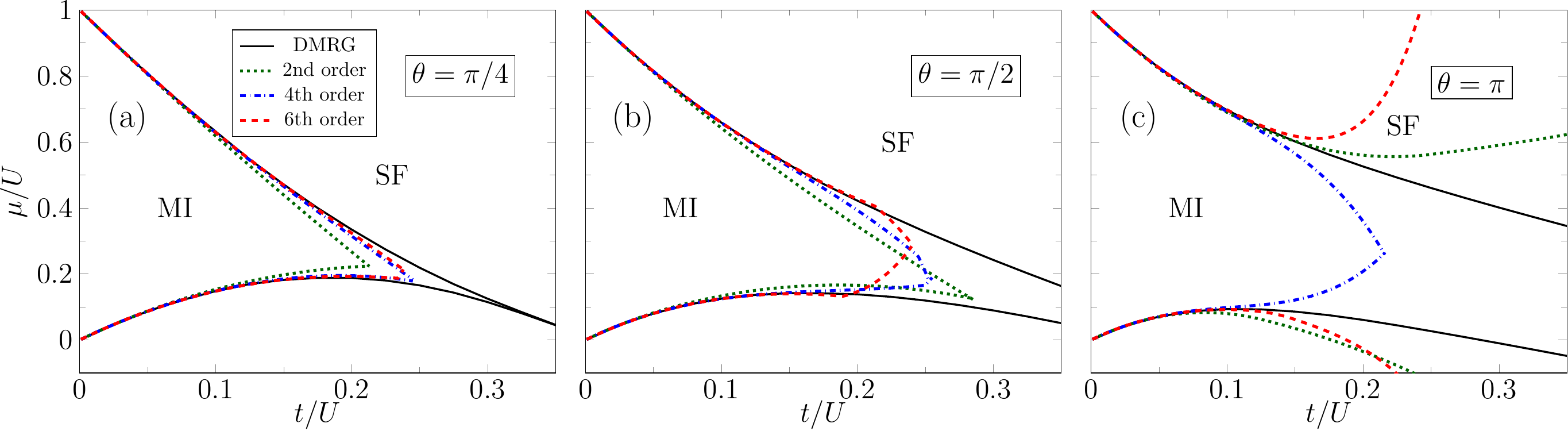}
 \end{center}
 \caption{(Color online)
 Phase diagram of the one-dimensional anyon-Hubbard model 
 ($n_{\rm p}\leq5$) for the fractional angle $\theta=\pi/4$ [panel (a)],
 $\pi/2$ [panel (b)], and $\pi$ [panel (c)] with  Mott insulator (MI) and superfluid
 (SF) regions. The MI-SF boundaries (black lines) 
 were determined by DMRG with system sizes up to $L=128$,
 open boundary conditions, and $n_{\rm p}\leq5$. The strong-coupling 
 expansions up to 6th order in $x$ show reasonable 
 agreements with the numerical data. 
 }
 \label{fig:pd}
\end{figure*}

As in the case of the BHM~\cite{KWM00,EFG11}, $\mu^\pm$ in the AHM 
can be also determined numerically by DMRG using the following definitions 
of the chemical potentials for finite system sizes
\begin{equation}
 \pm \mu^\pm(L)=E_0(L,N\pm 1)-E_0(L,N)\, ,
\end{equation}
where $E_0(L,N\pm 1)$ and $E_0(L,N)$ denotes the corresponding ground-state energies. 

Figure~\ref{fig:pd} shows the ground-state phase diagram of the 1D 
AHM, exhibiting MI and SF regions as a function of the chemical
potential $\mu/U$ and the anyon transfer amplitude $t/U$.
The strong-coupling expansions of the chemical potentials via 
Eqs.~\eqref{Eh} and \eqref{Ep} up to 6th order are compared 
with DMRG results. For small $\theta\lesssim\pi/2$, both methods
essentially agree up to $x\lesssim 0.2$ 
[see Figs.~\ref{fig:pd}(a) and (b)], while in the case of the pseudo\-fermions [$\theta=\pi$ in Fig.~\ref{fig:pd}(c)] 
even 6th-order results start to deviate around $x\sim 0.12$. 
In the intermediate-coupling regime
($t/U\gtrsim 0.20$) sudden changes will appear in the perturbation
results, especially for $\theta=\pi/4$ (not shown).
This is because the perturbation expansions fall into
the wrong minima as will be discussed in App.~\ref{sec:minima}.

\subsection{Momentum distribution function}
Anyons might be characterized most significantly by momentum
distribution functions as has been demonstrated for both hardcore~\cite{HZC08}
and softcore~\cite{TEP15} anyons. 
For the current model, we can define two different types of single-particle correlation functions 
\begin{eqnarray}
 C_b(r)&=&\langle \hat{b}_{j}^\dagger \hat{b}_{j+r}^{\phantom{\dagger}}\rangle\, ,
  \\
 C_a(r)&=&\langle \hat{a}_{j}^\dagger \hat{a}_{j+r}^{\phantom{\dagger}}\rangle\, ,
\end{eqnarray}
corresponding to boson or anyon representations. 
Results for the boson correlation function should be relevant for the proposed realization of the model in optical lattices. 
The anyon correlation function can be expressed in terms of the boson operators as follows: 
\begin{equation}
 \langle\hat{a}_j^{\dagger}\hat{a}_\ell^{\phantom{\dagger}}\rangle
 \to \left\{ \begin{array}{ll}

   \left\langle\hat{b}_j^{\dagger}e^{i\theta\hat{n}_j}
    \left[\displaystyle\prod_{j<m<\ell}
     e^{i\theta\hat{n}_m}
    \right]
    \hat{b}_\ell^{\phantom{\dagger}} \right\rangle
    & {\rm for\ } j<\ell\,, \\
   \left\langle 
    e^{-i\theta\hat{n}_\ell}\hat{b}_\ell^{\phantom{\dagger}}
    \left[\displaystyle\prod_{\ell<m<j}
     e^{-i\theta\hat{n}_m}
    \right]
    \hat{b}_j^{\dagger} 
   \right\rangle

    & {\rm for\ } j>\ell\,,\\
      \langle\hat{n}_j\rangle & {\rm for\ } j=\ell \,.\\
     \end{array} \right.
\label{anyon-corr}
\end{equation}

Within the strong-coupling expansion, the above correlators 
are translated according to   
\begin{eqnarray}
  \langle\hat{b}_j^{\dagger}\hat{b}_\ell^{\phantom{\dagger}}\rangle
   &\mapsto& 
   \langle \phi_0|
   e^{\hat{S}}
   \hat{b}_j^{\dagger}\hat{b}_\ell^{\phantom{\dagger}}
   e^{-\hat{S}}
   |\phi_0\rangle\, ,
    \\
   \langle\hat{a}_j^{\dagger}\hat{a}_\ell^{\phantom{\dagger}}\rangle
    &\mapsto&
   \langle \phi_0|
   e^{\hat{S}}
   \hat{a}_j^{\dagger}\hat{a}_\ell^{\phantom{\dagger}}
   e^{-\hat{S}}
   |\phi_0\rangle\, .
   \label{ab-correlator-sc}
\end{eqnarray}
In App.~\ref{sec:corr-func} we compare the perturbation results for the two-point 
correlation functions up to 4th order in $x$ with the iDMRG data. 

The Fourier-transformed single-particle density matrices
give the momentum distribution functions for bosons and anyons as
\begin{eqnarray}
 n_b(k)=\frac{1}{L}\sum_{j,\ell}e^{ik(j-\ell)}
  \langle 
  \hat{b}_j^{\dagger}\hat{b}_\ell^{\phantom{\dagger}}
  \rangle\,,
  \label{nbk}
  \\
 n_a(k)=\frac{1}{L}\sum_{j,\ell}e^{ik(j-\ell)}
  \langle 
  \hat{a}_j^{\dagger}\hat{a}_\ell^{\phantom{\dagger}}
  \rangle\,.
  \label{nak}
\end{eqnarray}
Up to and including 4th order in $x$, we obtain for the
momentum distribution functions of bosons 
\begin{eqnarray}
 n_b^{[4]}(k)&=&1+x\{4 \cos{(k )} + 4 \cos{(k + \theta )}\}
\nonumber \\ 
 && +x^{2}\{4 \cos{(2 k )} + 24 \cos{(2 k + \theta )} 
\nonumber \\ 
 && \phantom{+x^{2}\{}
  + 8 \cos{(2 k + 2 \theta )}\}  
\nonumber \\ 
&& +x^{3} \{- 36 \cos{(k )} + 12 \cos{(k - \theta )} 
\nonumber \\ 
 && \phantom{+x^{3}\{}
 - 8 \cos{(k + \theta )} + 16 \cos{(k + 2 \theta )} 
\nonumber \\ 
 && \phantom{+x^{3}\{}
 + 4 \cos{(3 k )} + 60 \cos{(3 k + \theta )} 
\nonumber \\ 
 && \phantom{+x^{3}\{}
 + 96 \cos{(3 k + 2 \theta )} + 16 \cos{(3 k + 3 \theta )}
  \}  
\nonumber \\ 
&& +x^{4} \biggl\{
 \frac{152}{3} \cos{\left (2 k \right )} 
 + \frac{64}{3} \cos{\left (2 k - \theta \right )} 
\nonumber \\ 
 && \phantom{+x^{4}\{}
 - 400 \cos{\left (2 k + \theta \right )} 
 + \frac{152}{3} \cos{\left (2 k + 2 \theta \right )} 
\nonumber \\ 
 && \phantom{+x^{4}\{}
 + 64 \cos{\left (2 k + 3 \theta \right )}
  +4 \cos{(4 k )} 
\nonumber \\ 
 && \phantom{+x^{4}\{}
+ 112 \cos{(4 k + \theta )} 
 + 432 \cos{(4 k + 2 \theta )} 
\nonumber \\ 
 && \phantom{+x^{4}\{}
 + 320 \cos{(4 k + 3 \theta )} 
 + 32 \cos{(4 k + 4 \theta )}
 \biggr\}\,,  \nonumber \\
\label{eq-nkb}
\end{eqnarray}
and for anyons
\begin{eqnarray}
n_a^{[4]}(k)&=& 1+8 x \cos{(k)} 
 \nonumber\\
&&+ x^{2}\{
           24 \cos{(2k)} 
	   + 12\cos{(2 k - \theta)}
         \}
 \nonumber\\
&& + x^{3} \{- 96 \cos{(k)} 
             + 40 \cos{(k + \theta)} 
  \nonumber\\
&& \phantom{+x^3\{ } 
             + 40 \cos{(k - \theta)} 
	     + 64 \cos{(3 k)} 
	       \nonumber\\
&& \phantom{+x^3\{ } 
           + 16 \cos{(3 k - 2 \theta )} 
           + 96 \cos{(3 k - \theta)}
	   \}  
  \nonumber\\
&& + x^{4} \biggl\{
 - \frac{1408}{3} \cos{(2 k  )} 
 + \frac{272}{3} \cos{(2 k - 2 \theta  )} 
   \nonumber\\
&& \phantom{+x^4\{ } 
 - \frac{176}{3} \cos{(2 k - \theta  )} 
 + 224 \cos{(2 k + \theta  )}
   \nonumber\\
&& \phantom{+x^4\{ } 
 + 160 \cos{(4 k)} + 480 \cos{(4 k - \theta)}
   \nonumber\\
&& \phantom{+x^4\{ } 
             + 240 \cos{(4 k - 2 \theta )} 
	     + 20 \cos{(4 k - 3 \theta)} 
   \biggr\}\, . 
	   \nonumber\\
\label{eq-nka}
\end{eqnarray}
Taking the limit $\theta\to0$ in Eqs.~\eqref{eq-nkb} or \eqref{eq-nka}, 
we obtain the momentum distribution function in the BHM
\begin{eqnarray}
 n_{\rm BHM}^{[4]}(k)&=&1+8x\cos(k)+36x^2\cos(2k)
  \nonumber\\
 && +x^3\{-16\cos(k)+176\cos(3k)\}
  \nonumber\\
 && +x^4\left\{-\frac{640}{3}\cos(2k)+900\cos(4k)\right\},
\end{eqnarray}
in agreement with the former studies of the strong-coupling
expansions in the BHM up to and including third order 
in $x$~\cite{FKKKT09}.

\begin{figure*}[t!]
 \begin{center}
  \includegraphics[width=0.85\linewidth,clip]{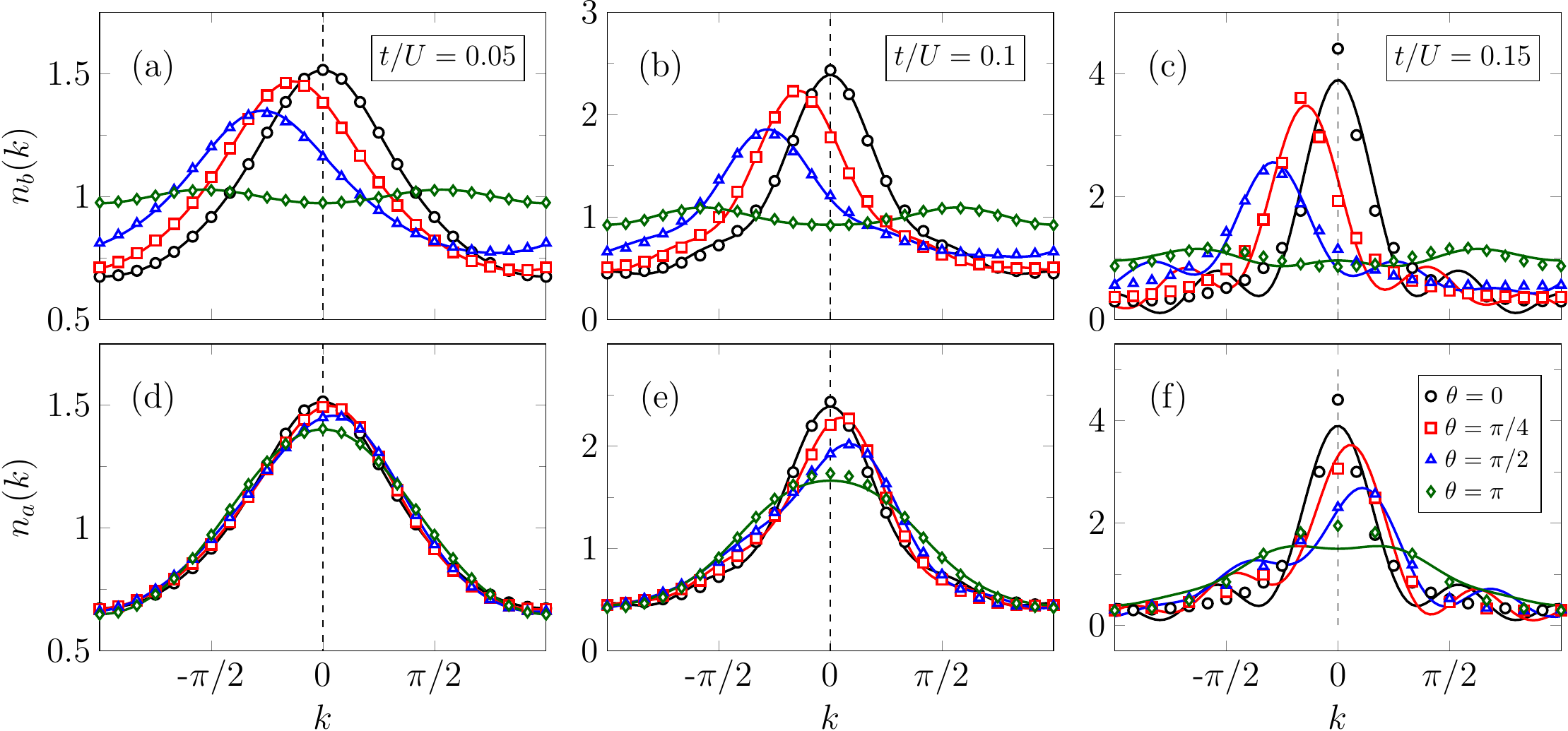}
 \end{center}
 \caption{(Color online) Momentum distribution function $n_b(k)$
 (upper panels) and $n_a(k)$ (lower panels) within the first Mott
 for various $\theta$ from DMRG with $L=48$ and PBCs (symbols)
 compared with 4th-order strong-coupling expansions (solid lines).
 }
 \label{fig:nk}
\end{figure*}

Using DMRG with PBCs, the momentum distribution 
functions of anyons and bosons can be extracted from Eqs.~\eqref{nbk}
and \eqref{nak} after calculating the two-point correlation functions,
as demonstrated in Fig.~\ref{fig:nk} by the comparison with the strong-coupling
expansions~\eqref{eq-nkb} and \eqref{eq-nka}. While for $t/U=0.05$ 
analytical and numerical methods agree [Fig.~\ref{fig:nk}(a)], small 
deviations appear for $t/U\gtrsim0.1$ [Fig.~\ref{fig:nk}(b)]. 
For $t/U\sim0.20$ [Fig.~\ref{fig:nk}(c)] the oscillations become 
significant in the 4th-order strong-coupling expansions
which are clearly an artifact. 

Analogous to the momentum-dependent excitation energies 
in Sec.~\ref{sec:ep-eh},
the characteristic asymmetry in the momentum
distribution functions can be understood by considering the 
main $\theta$-dependent contributions in the strong-coupling expansion 
of $n_{b/a}(k)$. 
In the BHM limit ($\theta=0$), $n_b(k) [=n_a(k)]$ is always symmetric
about $k=0$, where the position of the maximum is located. 
These peak positions of $n_b(k)$ [$n_a(k)$] shift to the negative 
[positive] momentum with increasing $\theta$ for $0<\theta<\pi$, 
which is consistent with the sign of $\theta$ in the cosine term of the 
main $\theta$-dependent contribution of $n_{b/a}(k)$, 
i.e., the positive [negative]
sign of $\theta$ in $n_b^{[1]}(k)$ [$n_a^{[2]}(k)$] of
Eq.~\eqref{eq-nkb} [Eq.~\eqref{eq-nka}]. Moreover, the peak positions 
of $n_b(k)$ depend more strongly on $\theta$ than those of $n_a(k)$,
this is because the $\theta$-dependent main contribution in $n_a(k)$
shows up first in the 2nd-order expansion, while in the case of $n_b(k)$
it can already be seen in the 1st-order expansion.  
To 1st order, the peak of $n_b(k)$ is located at $k=-\theta/2$, that is, its position depends linearly on the fractional angle. 
When increasing $\theta$ from $0$ to $\pi$, the boson momentum distribution becomes flatter because of cancellations in the 1st-order terms. Close to the pseudofermion limit, a second peak appears for $0<k<\pi$ that can be attributed mainly to 2nd-order contributions. At $\theta=\pi$, the cancellation of 1st-order terms becomes exact for all $k$ and one ends up with $n_b(k)=1-12x^2\cos(2k)+{\cal O}(x^3)$. 
Our results for the boson momentum distribution function in the MI are in contrast to the results of Ref.~\cite{TEP15} for the SF where, depending on the filling, a single peak at either $k\sim 0$ or $k\sim -\theta$ has been found.

\section{Summary and outlook}
\label{sec:summary}
We studied the MI phase of the anyon-Hubbard model at filling factor one  for arbitrary fractional angle $\theta$ using strong-coupling perturbation theory. 
Explicit expressions for the $\theta$-dependence of the ground state energy per site, the single-particle and single-hole excitation energies as well as the momentum distribution functions were obtained for up to 6th order in $t/U$ (hopping/interaction).

In the BHM both single-particle and single-hole dispersions have their minimum at $k=0$. 
For finite $\theta$, the minimum of the dispersion is shifted differently for single-particle and single-hole excitations, that is, there is an indirect gap for particle-hole excitations. 

The momentum distribution functions become asymmetric for $0<\theta<\pi$ with the peak shifted to negative (positive) momentum in the boson (anyon) description of the model. 
A stronger $\theta$-dependence is found for the boson momentum distribution than for the anyon one. In particular, the boson momentum distribution function becomes almost flat in the pseudofermion limit $\theta=\pi$.

While the series generated by the strong-coupling expansion might be asymptotic, the results for finite order agree well with numerically exact MPS calculations for small hopping $t$.  
Increasing $t$, however, the accuracy starts to deteriorate and the perturbative description is no longer sensible. 
At fixed order, the region of validity of the perturbative expansion seemingly decreases when the fractional angle $\theta$ is increased even though the MI region becomes larger. 

%
Obviously, there are several other directions to extend our work. 
The perturbation results in this paper are limited to the Mott 
insulator for $\rho=1$. First natural extension might be the 
strong-coupling study in higher integer fillings.  Another possibility is the inclusion of a nearest-neighbor interaction which leads to additional Haldane-insulator and density-wave phases~\cite{LEF17}, the latter being susceptible to a perturbative treatment.  
Furthermore, the strong-coupling approach could be applied to the AHM in higher dimensions. In this case perturbation theory should be particularly useful since no quasi-exact results from MPS-based methods are available. 
Finally, for the comparison with future experiments it would be desirable 
to investigate the dynamical quantities, such as the single-particle 
spectral functions, the dynamical structure factor, and 
the dynamical current and kinetic-energy correlation functions 
as demonstrated in the BHM~\cite{EFG12,EFGzMKAvdL12,zMGEF14}.

\section*{Acknowledgments}We thank F. Gebhard for valuable discussions.    
DMRG simulations were performed using the ITensor 
library~\cite{ITensor}. This work was supported
by Deutsche Forschungsgemeinschaft (Germany) through SFB 652.

\appendix

\section{Excitation energies in the Bose-Hubbard limit}
\label{sec:BH-limit}
Taking the limit $\theta=0$ in Eqs.~\eqref{Eh} and \eqref{Ep}
we obtain the single-hole and single-particle excitation 
energies in the momentum space for the BHM as 
\begin{eqnarray}
 \frac{E_{\rm h}(k)}{t} &=& 8x -\frac{512}{3}x^{5} 
  \nonumber\\
&&+\left(-2+12x^{2}-\frac{224}{3}x^{4}\right)\cos (k) 
 \nonumber\\
&&+\left(-4x+64x^{3}-\frac{1436}{3}x^{5}\right)\cos (2k)
 \nonumber \\
&&+\left(-12x^{2}+276x^{4}\right)\cos (3k)  \nonumber\\
&&+\left(-44x^{3}+1296x^{5}\right)\cos (4k) \nonumber\\
&&-180x^{4}\cos (5k) -788 x^{5}\cos (6k) 
\nonumber\\
&& + {\cal O}\left({x^{6}}\right) \; ,
 \label{Eh-BHM} 
\end{eqnarray}
and 
\begin{eqnarray}
\frac{E_{\rm p}(k)}{t} &=& 
\frac{1}{x}+5x-\frac{513}{20}x^{3}-\frac{80139}{200}x^{5} \nonumber\\
&&+\left(-4+18x^{2}-\frac{137}{150}x^{4}\right)\cos (k) \nonumber\\
&&+\left(-4x+64x^{3}-\frac{426161}{750}x^{5}\right)\cos (2k) \nonumber 
\\
&&+\left(-12x^{2}+276x^{4}\right)\cos (3k)  \nonumber\\
&&+\left(-44x^{3}+1296x^{5}\right)\cos (4k) \nonumber\\
&&-180x^{4}\cos (5k) -788x^{5}\cos (6k) 
\nonumber\\
&&+ {\cal O}\left({x^{6}}\right) \; .
\label{Ep-BHM}
\end{eqnarray}
Direct comparison of Eqs.~\eqref{Eh-BHM} and \eqref{Ep-BHM} 
with iMPS results is demonstrated in Fig.~\ref{gap-k}.
Note that in Eqs.~(24) and (25) of Ref.~\cite{EFGzMKAvdL12}
the coefficient of the $x^5\cos(6k)$ term in $E_{\rm h}(k)$,
as well as the coefficients of the $x^5\cos(2k)$ and the $x^5\cos(6k)$ terms in $E_{\rm p}(k)$
are flawed; these errors were corrected in Eqs.~\eqref{Eh-BHM} and \eqref{Ep-BHM}.

Plotting chemical potentials, $\mu_{+}=\min\{E_{\rm p}(k)\}$ and 
$\mu_{-}=-\min\{E_{\rm h}(k)\}$, the phase diagram of the BHM can be
extracted from the strong-coupling expansions, Eqs.~\eqref{Eh-BHM} and
\eqref{Ep-BHM},  and compared  with the DMRG prediction (see Fig.~\ref{pd-bhm}).

\begin{figure}[tb]
 \begin{center}
  \includegraphics[width=0.85\linewidth,clip]{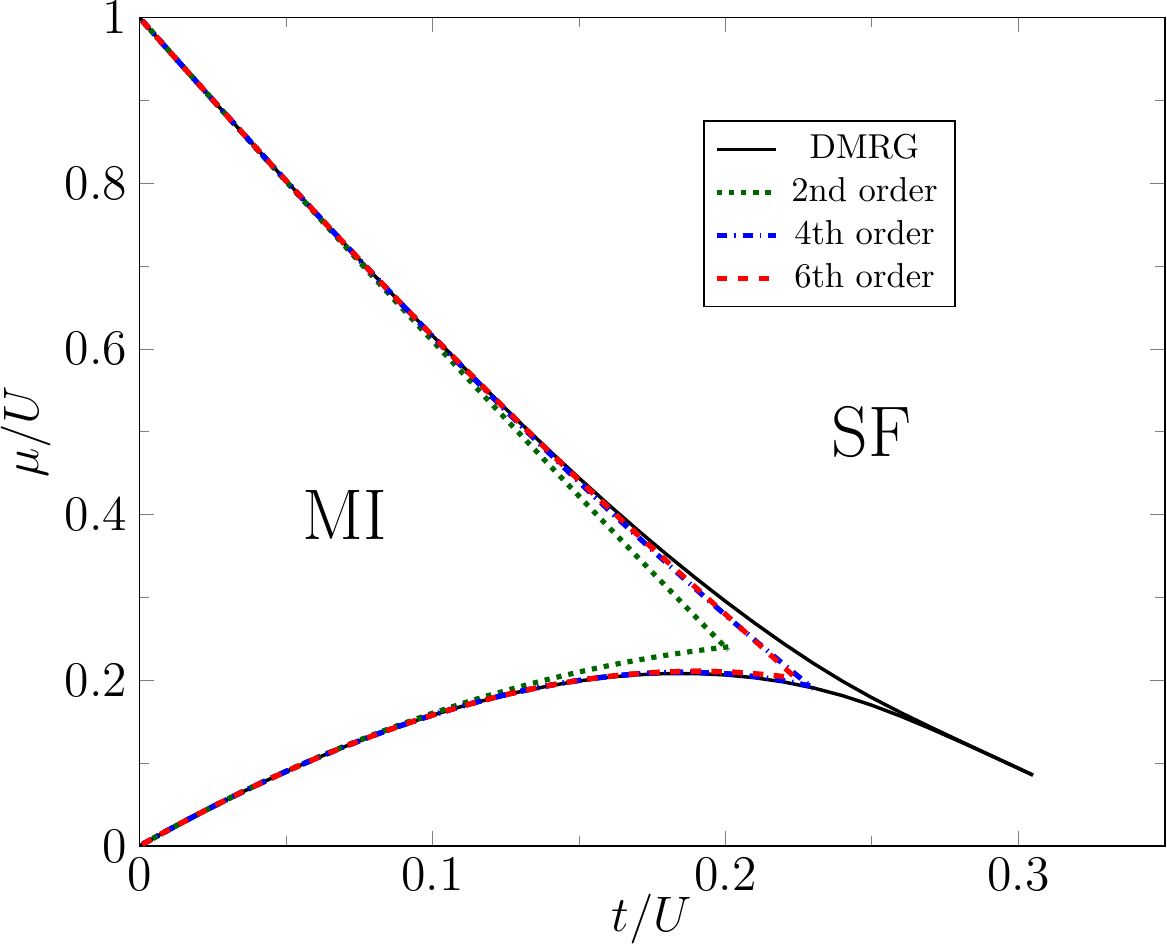}
 \end{center}
 \caption{(Color online) Ground-state phase diagram of the Bose--Hubbard
 model ($\theta=0$). The strong-coupling results \eqref{Eh-BHM} and 
 \eqref{Ep-BHM} show a reasonable agreement  
 with the DMRG data (black solid line). 
 }
 \label{pd-bhm}
\end{figure}
 
\section{Minima of excitation energies}
\label{sec:minima}
In Fig.~\ref{fig:pd}, chemical potentials $\mu^{\pm}$, 
obtained from strong-coupling expansions, show a sudden increase
for large $t/U (\gtrsim 0.24)$, especially for $\theta=\pi/4$ 
[see, e.g., the results of Fig.~\ref{gap-k-quarter-t025}(a), which were not included
in Fig.~\ref{fig:pd}.]. 
In this section we explain the origin of this shortcoming in detail. 

Figure~\ref{gap-k-quarter-t025}(b) shows the single-hole excitation
energies by the 6th-order strong-coupling expansion~\eqref{Eh} 
for $t/U=0.25$ and $\theta=\pi/4$ compared with iMPS results.
In this intermediate-coupling region the perturbation results 
oscillate strongly, so that the position of the minimum for 
$E_{\rm h}^{[6]}(k)$ to estimate $\mu^-=\min\{E_{\rm h}^{[6]}(k)\}$
shift from negative (star symbol) to positive (cross symbol)
momentum, while iMPS data still indicate that the minimum should be
located at the negative momentum as in the case of $t/U<0.25$.  
This sudden change of the location of minima leads to the artificial 
upturns of the strong-coupling expansions in the intermediate-coupling
region of Fig.~\ref{fig:pd}.

\begin{figure}[tb]
 \begin{center}
  \includegraphics[width=0.85\linewidth,clip]{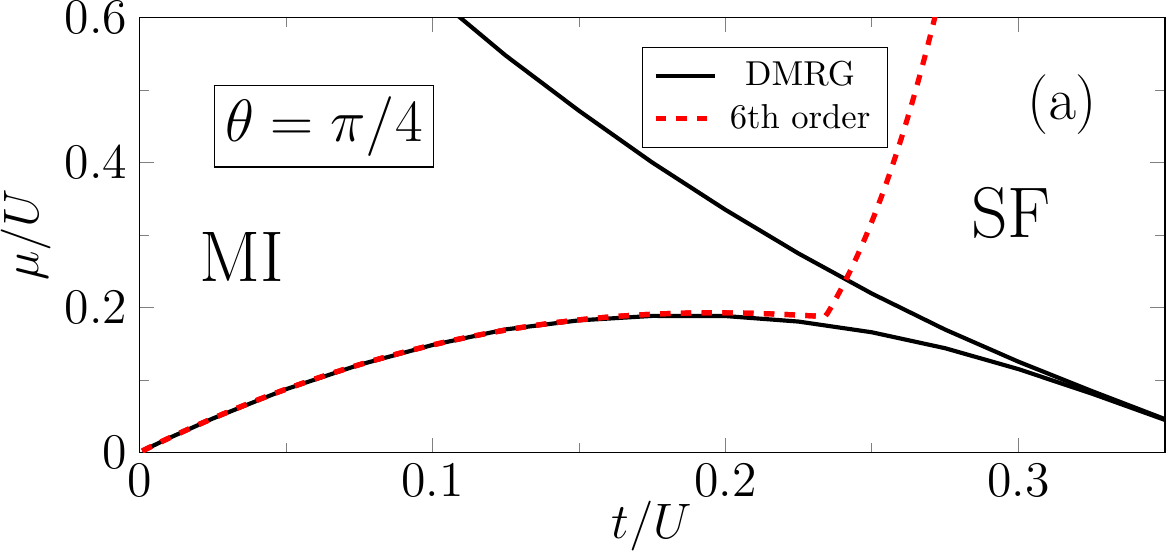}
  \includegraphics[width=0.85\linewidth,clip]{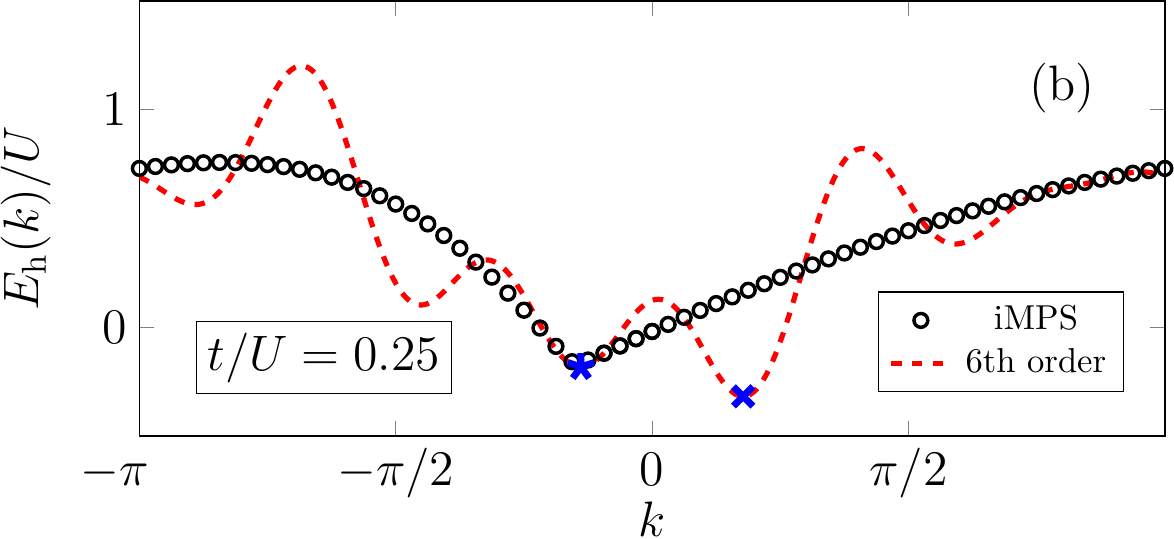}
 \end{center}
 \caption{(Color online) (a): Zoomed view of Fig.~\ref{fig:pd}(a)
showing the artificial upturn of the 6th-order 
 perturbation result $\mu^-$ at $\theta=\pi/4$.
 (b): Single-hole excitation energy $E_{\rm h}(k)$
 in the momentum space for $\theta=\pi/4$ and $t/U=0.25$ 
 by the 6th-order strong-coupling expansion (dashed line) 
 compared with the iMPS results (circles). 
 Star and cross symbols denote the correct and wrong minima 
 to estimate chemical potential $\mu^-$, see text.
 }
 \label{gap-k-quarter-t025}
\end{figure}

\section{Correlation function}
\label{sec:corr-func}
In this section we will give  the strong-coupling expressions for
the boson and anyon correlation functions.
Note that in the anyonic case Eq.~\eqref{anyon-corr} should be taken
into account.
As explained in the main text, we employ the Harris-Lange transformation
and obtain $C_{b/a}(r)$ for the distance $r=1$ to 4 up to 4th order 
in $x=t/U$ as
\begin{eqnarray}
 C_b(1)&=& x(2+2e^{-\itheta})
  \nonumber\\
 \label{eq-Cb1}
 && +x^3(-18+6e^{\itheta}-4e^{-\itheta}
         +8e^{-2\itheta})\,,
 \\
 C_b(2)&=& x^2(2+12e^{-\itheta}+4e^{-2\itheta})
  \\
 &&+x^4\left(
	\frac{76}{3}+\frac{32}{3}e^{\itheta}
	-200e^{-\itheta}+\frac{76}{3}e^{-2\itheta}
       \right)\, ,
 \nonumber\\
 C_b(3)&=& x^3(2+30e^{-\itheta}+48e^{-2\itheta}
           +8e^{-3\itheta})\,,
  \\
 C_b(4)&=& x^4(56e^{-\itheta}+216e^{-2\itheta}
          +160e^{-3\itheta}+16e^{-4\itheta})\, ,
  \nonumber \\
\label{eq-Cb4}
\end{eqnarray}
and 
\begin{eqnarray}
 C_a(1)&=& 4x+x^3(-48+20e^{-\itheta}+20e^{\itheta})\, ,
  \label{eq-Ca1}
 \\
 C_a(2)&=& x^2(12+6e^{\itheta})
  \\
 && +x^4\left(
       -\frac{704}{3}+\frac{136}{3}e^{2\itheta}
       -\frac{88}{3}e^{\itheta}+112e^{-\itheta}
      \right)\, ,
 \nonumber\\
 C_a(3) &=& x^3(32+8e^{2\itheta}+48e^{\itheta})\, ,
  \\
 C_a(4) &=& x^4(80+240e^{\itheta}+120e^{2\itheta}
  +10e^{3\itheta})\,.
\label{eq-Ca4}
\end{eqnarray}
Finally, the Fourier-transforms of Eqs.~\eqref{eq-Cb1}-\eqref{eq-Cb4} 
[\eqref{eq-Ca1}-\eqref{eq-Ca4}] provide us with 
the momentum distribution function~\eqref{eq-nkb} [\eqref{eq-nka}].

These two-point functions can be compared directly with the 
iDMRG data for $\chi=100$. Figure~\ref{corr-func} demonstrates for the real part
that strong-coupling and iDMRG results are in excellent agreement for all fractional angles 
$\theta$ at $x=0.05$. 

\begin{figure}[t!]
 \begin{center}
  \includegraphics[width=0.85\linewidth,clip]{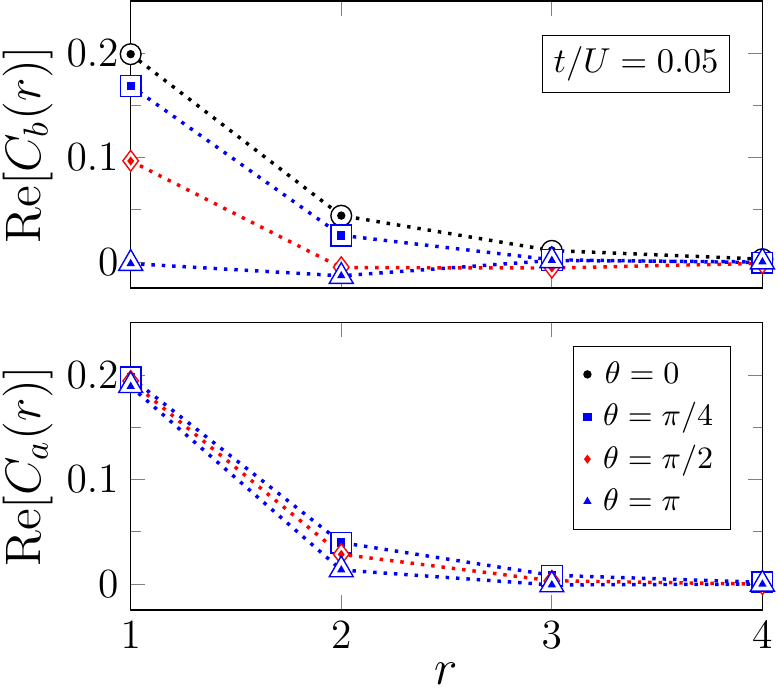}
 \end{center}
 \caption{(Color online) Strong-coupling results of the 
 boson (upper panel) and anyon (lower panel) correlation 
 functions $C_{b/a}(r)$ (closed symbols) for the distance 
 $r=1$ to 4, compared with the iDMRG data for $\chi=100$
 (open symbols).
 }
 \label{corr-func}
\end{figure}

\bibliography{ref-ahm}

\begin{thebibliography}{29}%
\makeatletter
\providecommand \@ifxundefined [1]{%
 \@ifx{#1\undefined}
}%
\providecommand \@ifnum [1]{%
 \ifnum #1\expandafter \@firstoftwo
 \else \expandafter \@secondoftwo
 \fi
}%
\providecommand \@ifx [1]{%
 \ifx #1\expandafter \@firstoftwo
 \else \expandafter \@secondoftwo
 \fi
}%
\providecommand \natexlab [1]{#1}%
\providecommand \enquote  [1]{``#1''}%
\providecommand \bibnamefont  [1]{#1}%
\providecommand \bibfnamefont [1]{#1}%
\providecommand \citenamefont [1]{#1}%
\providecommand \href@noop [0]{\@secondoftwo}%
\providecommand \href [0]{\begingroup \@sanitize@url \@href}%
\providecommand \@href[1]{\@@startlink{#1}\@@href}%
\providecommand \@@href[1]{\endgroup#1\@@endlink}%
\providecommand \@sanitize@url [0]{\catcode `\\12\catcode `\$12\catcode
  `\&12\catcode `\#12\catcode `\^12\catcode `\_12\catcode `\%12\relax}%
\providecommand \@@startlink[1]{}%
\providecommand \@@endlink[0]{}%
\providecommand \url  [0]{\begingroup\@sanitize@url \@url }%
\providecommand \@url [1]{\endgroup\@href {#1}{\urlprefix }}%
\providecommand \urlprefix  [0]{URL }%
\providecommand \Eprint [0]{\href }%
\providecommand \doibase [0]{http://dx.doi.org/}%
\providecommand \selectlanguage [0]{\@gobble}%
\providecommand \bibinfo  [0]{\@secondoftwo}%
\providecommand \bibfield  [0]{\@secondoftwo}%
\providecommand \translation [1]{[#1]}%
\providecommand \BibitemOpen [0]{}%
\providecommand \bibitemStop [0]{}%
\providecommand \bibitemNoStop [0]{.\EOS\space}%
\providecommand \EOS [0]{\spacefactor3000\relax}%
\providecommand \BibitemShut  [1]{\csname bibitem#1\endcsname}%
\let\auto@bib@innerbib\@empty
\bibitem [{\citenamefont {Leinaas}\ and\ \citenamefont {Myrheim}(1977)}]{LM77}%
  \BibitemOpen
  \bibfield  {author} {\bibinfo {author} {\bibfnamefont {J.~M.}\ \bibnamefont
  {Leinaas}}\ and\ \bibinfo {author} {\bibfnamefont {J.}~\bibnamefont
  {Myrheim}},\ }\href@noop {} {\bibfield  {journal} {\bibinfo  {journal} {Nuovo
  Cimento B}\ }\textbf {\bibinfo {volume} {37}},\ \bibinfo {pages} {1}
  (\bibinfo {year} {1977})}\BibitemShut {NoStop}%
\bibitem [{\citenamefont {Wilczek}(1982)}]{Wi82}%
  \BibitemOpen
  \bibfield  {author} {\bibinfo {author} {\bibfnamefont {F.}~\bibnamefont
  {Wilczek}},\ }\href {\doibase 10.1103/PhysRevLett.49.957} {\bibfield
  {journal} {\bibinfo  {journal} {Phys. Rev. Lett.}\ }\textbf {\bibinfo
  {volume} {49}},\ \bibinfo {pages} {957} (\bibinfo {year} {1982})}\BibitemShut
  {NoStop}%
\bibitem [{\citenamefont {Tsui}\ \emph {et~al.}(1982)\citenamefont {Tsui},
  \citenamefont {Stormer},\ and\ \citenamefont {Gossard}}]{TSG82}%
  \BibitemOpen
  \bibfield  {author} {\bibinfo {author} {\bibfnamefont {D.~C.}\ \bibnamefont
  {Tsui}}, \bibinfo {author} {\bibfnamefont {H.~L.}\ \bibnamefont {Stormer}}, \
  and\ \bibinfo {author} {\bibfnamefont {A.~C.}\ \bibnamefont {Gossard}},\
  }\href {\doibase 10.1103/PhysRevLett.48.1559} {\bibfield  {journal} {\bibinfo
   {journal} {Phys. Rev. Lett.}\ }\textbf {\bibinfo {volume} {48}},\ \bibinfo
  {pages} {1559} (\bibinfo {year} {1982})}\BibitemShut {NoStop}%
\bibitem [{\citenamefont {Laughlin}(1983)}]{Laughlin83}%
  \BibitemOpen
  \bibfield  {author} {\bibinfo {author} {\bibfnamefont {R.~B.}\ \bibnamefont
  {Laughlin}},\ }\href {\doibase 10.1103/PhysRevLett.50.1395} {\bibfield
  {journal} {\bibinfo  {journal} {Phys. Rev. Lett.}\ }\textbf {\bibinfo
  {volume} {50}},\ \bibinfo {pages} {1395} (\bibinfo {year}
  {1983})}\BibitemShut {NoStop}%
\bibitem [{\citenamefont {Haldane}(1991)}]{Haldane91}%
  \BibitemOpen
  \bibfield  {author} {\bibinfo {author} {\bibfnamefont {F.~D.~M.}\
  \bibnamefont {Haldane}},\ }\href {\doibase 10.1103/PhysRevLett.67.937}
  {\bibfield  {journal} {\bibinfo  {journal} {Phys. Rev. Lett.}\ }\textbf
  {\bibinfo {volume} {67}},\ \bibinfo {pages} {937} (\bibinfo {year}
  {1991})}\BibitemShut {NoStop}%
\bibitem [{\citenamefont {Keilmann}\ \emph {et~al.}(2011)\citenamefont
  {Keilmann}, \citenamefont {Lanzmich}, \citenamefont {McCulloch},\ and\
  \citenamefont {Roncaglia}}]{KLMR11}%
  \BibitemOpen
  \bibfield  {author} {\bibinfo {author} {\bibfnamefont {T.}~\bibnamefont
  {Keilmann}}, \bibinfo {author} {\bibfnamefont {S.}~\bibnamefont {Lanzmich}},
  \bibinfo {author} {\bibfnamefont {I.}~\bibnamefont {McCulloch}}, \ and\
  \bibinfo {author} {\bibfnamefont {M.}~\bibnamefont {Roncaglia}},\ }\href
  {\doibase 10.1038/ncomms1353} {\bibfield  {journal} {\bibinfo  {journal}
  {Nat. Commun.}\ }\textbf {\bibinfo {volume} {2}},\ \bibinfo {pages} {361}
  (\bibinfo {year} {2011})}\BibitemShut {NoStop}%
\bibitem [{\citenamefont {Greschner}\ and\ \citenamefont
  {Santos}(2015)}]{GS15}%
  \BibitemOpen
  \bibfield  {author} {\bibinfo {author} {\bibfnamefont {S.}~\bibnamefont
  {Greschner}}\ and\ \bibinfo {author} {\bibfnamefont {L.}~\bibnamefont
  {Santos}},\ }\href {\doibase 10.1103/PhysRevLett.115.053002} {\bibfield
  {journal} {\bibinfo  {journal} {Phys. Rev. Lett.}\ }\textbf {\bibinfo
  {volume} {115}},\ \bibinfo {pages} {053002} (\bibinfo {year}
  {2015})}\BibitemShut {NoStop}%
\bibitem [{\citenamefont {Str\"ater}\ \emph {et~al.}(2016)\citenamefont
  {Str\"ater}, \citenamefont {Srivastava},\ and\ \citenamefont
  {Eckardt}}]{SSE16}%
  \BibitemOpen
  \bibfield  {author} {\bibinfo {author} {\bibfnamefont {C.}~\bibnamefont
  {Str\"ater}}, \bibinfo {author} {\bibfnamefont {S.~C.~L.}\ \bibnamefont
  {Srivastava}}, \ and\ \bibinfo {author} {\bibfnamefont {A.}~\bibnamefont
  {Eckardt}},\ }\href {\doibase 10.1103/PhysRevLett.117.205303} {\bibfield
  {journal} {\bibinfo  {journal} {Phys. Rev. Lett.}\ }\textbf {\bibinfo
  {volume} {117}},\ \bibinfo {pages} {205303} (\bibinfo {year}
  {2016})}\BibitemShut {NoStop}%
\bibitem [{\citenamefont {Tang}\ \emph {et~al.}(2015)\citenamefont {Tang},
  \citenamefont {Eggert},\ and\ \citenamefont {Pelster}}]{TEP15}%
  \BibitemOpen
  \bibfield  {author} {\bibinfo {author} {\bibfnamefont {G.}~\bibnamefont
  {Tang}}, \bibinfo {author} {\bibfnamefont {S.}~\bibnamefont {Eggert}}, \ and\
  \bibinfo {author} {\bibfnamefont {A.}~\bibnamefont {Pelster}},\ }\href
  {http://stacks.iop.org/1367-2630/17/i=12/a=123016} {\bibfield  {journal}
  {\bibinfo  {journal} {New J. Phys.}\ }\textbf {\bibinfo {volume} {17}},\
  \bibinfo {pages} {123016} (\bibinfo {year} {2015})}\BibitemShut {NoStop}%
\bibitem [{\citenamefont {Arcila-Forero}\ \emph {et~al.}(2016)\citenamefont
  {Arcila-Forero}, \citenamefont {Franco},\ and\ \citenamefont
  {Silva-Valencia}}]{AFS16}%
  \BibitemOpen
  \bibfield  {author} {\bibinfo {author} {\bibfnamefont {J.}~\bibnamefont
  {Arcila-Forero}}, \bibinfo {author} {\bibfnamefont {R.}~\bibnamefont
  {Franco}}, \ and\ \bibinfo {author} {\bibfnamefont {J.}~\bibnamefont
  {Silva-Valencia}},\ }\href {\doibase 10.1103/PhysRevA.94.013611} {\bibfield
  {journal} {\bibinfo  {journal} {Phys. Rev. A}\ }\textbf {\bibinfo {volume}
  {94}},\ \bibinfo {pages} {013611} (\bibinfo {year} {2016})}\BibitemShut
  {NoStop}%
\bibitem [{\citenamefont {Zhang}\ \emph {et~al.}()\citenamefont {Zhang},
  \citenamefont {Greschner}, \citenamefont {Fan}, \citenamefont {Scott},\ and\
  \citenamefont {Zhang}}]{ZGFSZ16}%
  \BibitemOpen
  \bibfield  {author} {\bibinfo {author} {\bibfnamefont {W.}~\bibnamefont
  {Zhang}}, \bibinfo {author} {\bibfnamefont {S.}~\bibnamefont {Greschner}},
  \bibinfo {author} {\bibfnamefont {E.}~\bibnamefont {Fan}}, \bibinfo {author}
  {\bibfnamefont {T.~C.}\ \bibnamefont {Scott}}, \ and\ \bibinfo {author}
  {\bibfnamefont {Y.}~\bibnamefont {Zhang}},\ }\href {\doibase 10.1103/PhysRevA.95.053614} {\bibfield
  {journal} {\bibinfo  {journal} {Phys. Rev. A}\ }\textbf {\bibinfo {volume}
  {95}},\ \bibinfo {pages} {053614} (\bibinfo {year} {2017})}\BibitemShut
  {NoStop}%
\bibitem [{\citenamefont {Ejima}\ \emph
  {et~al.}(2012{\natexlab{a}})\citenamefont {Ejima}, \citenamefont {Fehske},
  \citenamefont {Gebhard}, \citenamefont {zu~M\"unster}, \citenamefont {Knap},
  \citenamefont {Arrigoni},\ and\ \citenamefont {von~der
  Linden}}]{EFGzMKAvdL12}%
  \BibitemOpen
  \bibfield  {author} {\bibinfo {author} {\bibfnamefont {S.}~\bibnamefont
  {Ejima}}, \bibinfo {author} {\bibfnamefont {H.}~\bibnamefont {Fehske}},
  \bibinfo {author} {\bibfnamefont {F.}~\bibnamefont {Gebhard}}, \bibinfo
  {author} {\bibfnamefont {K.}~\bibnamefont {zu~M\"unster}}, \bibinfo {author}
  {\bibfnamefont {M.}~\bibnamefont {Knap}}, \bibinfo {author} {\bibfnamefont
  {E.}~\bibnamefont {Arrigoni}}, \ and\ \bibinfo {author} {\bibfnamefont
  {W.}~\bibnamefont {von~der Linden}},\ }\href {\doibase
  10.1103/PhysRevA.85.053644} {\bibfield  {journal} {\bibinfo  {journal} {Phys.
  Rev. A}\ }\textbf {\bibinfo {volume} {85}},\ \bibinfo {pages} {053644}
  (\bibinfo {year} {2012}{\natexlab{a}})}\BibitemShut {NoStop}%
\bibitem [{\citenamefont {Damski}\ and\ \citenamefont
  {Zakrzewski}(2006)}]{DZ06}%
  \BibitemOpen
  \bibfield  {author} {\bibinfo {author} {\bibfnamefont {B.}~\bibnamefont
  {Damski}}\ and\ \bibinfo {author} {\bibfnamefont {J.}~\bibnamefont
  {Zakrzewski}},\ }\href {\doibase 10.1103/PhysRevA.74.043609} {\bibfield
  {journal} {\bibinfo  {journal} {Phys. Rev. A}\ }\textbf {\bibinfo {volume}
  {74}},\ \bibinfo {pages} {043609} (\bibinfo {year} {2006})}\BibitemShut
  {NoStop}%
\bibitem [{\citenamefont {Elstner}\ and\ \citenamefont {Monien}()}]{EM99b}%
  \BibitemOpen
  \bibfield  {author} {\bibinfo {author} {\bibfnamefont {N.}~\bibnamefont
  {Elstner}}\ and\ \bibinfo {author} {\bibfnamefont {H.}~\bibnamefont
  {Monien}},\ }\href {http://arxiv.org/abs/cond-mat/9905367} {}\bibinfo {note}
  {{a}rXiv:cond-mat/9905367}\BibitemShut {NoStop}%
\bibitem [{\citenamefont {White}(1992)}]{White92}%
  \BibitemOpen
  \bibfield  {author} {\bibinfo {author} {\bibfnamefont {S.~R.}\ \bibnamefont
  {White}},\ }\href {\doibase 10.1103/PhysRevLett.69.2863} {\bibfield
  {journal} {\bibinfo  {journal} {Phys. Rev. Lett.}\ }\textbf {\bibinfo
  {volume} {69}},\ \bibinfo {pages} {2863} (\bibinfo {year}
  {1992})}\BibitemShut {NoStop}%
\bibitem [{\citenamefont {McCulloch}()}]{Mc08}%
  \BibitemOpen
  \bibfield  {author} {\bibinfo {author} {\bibfnamefont {I.~P.}\ \bibnamefont
  {McCulloch}},\ }\href {http://arxiv.org/abs/0804.2509} {}\bibinfo {note}
  {{a}rXiv:0804.2509}\BibitemShut {NoStop}%
\bibitem [{\citenamefont {Schollw\"{o}ck}(2011)}]{Sch11}%
  \BibitemOpen
  \bibfield  {author} {\bibinfo {author} {\bibfnamefont {U.}~\bibnamefont
  {Schollw\"{o}ck}},\ }\href {\doibase 10.1016/j.aop.2010.09.012} {\bibfield
  {journal} {\bibinfo  {journal} {Ann. Phys.}\ }\textbf {\bibinfo {volume}
  {326}},\ \bibinfo {pages} {96} (\bibinfo {year} {2011})}\BibitemShut
  {NoStop}%
\bibitem [{\citenamefont {Haegeman}\ \emph {et~al.}(2012)\citenamefont
  {Haegeman}, \citenamefont {Pirvu}, \citenamefont {Weir}, \citenamefont
  {Cirac}, \citenamefont {Osborne}, \citenamefont {Verschelde},\ and\
  \citenamefont {Verstraete}}]{HPWCOVV12}%
  \BibitemOpen
  \bibfield  {author} {\bibinfo {author} {\bibfnamefont {J.}~\bibnamefont
  {Haegeman}}, \bibinfo {author} {\bibfnamefont {B.}~\bibnamefont {Pirvu}},
  \bibinfo {author} {\bibfnamefont {D.~J.}\ \bibnamefont {Weir}}, \bibinfo
  {author} {\bibfnamefont {J.~I.}\ \bibnamefont {Cirac}}, \bibinfo {author}
  {\bibfnamefont {T.~J.}\ \bibnamefont {Osborne}}, \bibinfo {author}
  {\bibfnamefont {H.}~\bibnamefont {Verschelde}}, \ and\ \bibinfo {author}
  {\bibfnamefont {F.}~\bibnamefont {Verstraete}},\ }\href {\doibase
  10.1103/PhysRevB.85.100408} {\bibfield  {journal} {\bibinfo  {journal} {Phys.
  Rev. B}\ }\textbf {\bibinfo {volume} {85}},\ \bibinfo {pages} {100408}
  (\bibinfo {year} {2012})}\BibitemShut {NoStop}%
\bibitem [{\citenamefont {Haegeman}\ \emph {et~al.}(2013)\citenamefont
  {Haegeman}, \citenamefont {Osborne},\ and\ \citenamefont
  {Verstraete}}]{HOV13}%
  \BibitemOpen
  \bibfield  {author} {\bibinfo {author} {\bibfnamefont {J.}~\bibnamefont
  {Haegeman}}, \bibinfo {author} {\bibfnamefont {T.~J.}\ \bibnamefont
  {Osborne}}, \ and\ \bibinfo {author} {\bibfnamefont {F.}~\bibnamefont
  {Verstraete}},\ }\href {\doibase 10.1103/PhysRevB.88.075133} {\bibfield
  {journal} {\bibinfo  {journal} {Phys. Rev. B}\ }\textbf {\bibinfo {volume}
  {88}},\ \bibinfo {pages} {075133} (\bibinfo {year} {2013})}\BibitemShut
  {NoStop}%
\bibitem [{\citenamefont {Harris}\ and\ \citenamefont {Lange}(1967)}]{HL67}%
  \BibitemOpen
  \bibfield  {author} {\bibinfo {author} {\bibfnamefont {A.~B.}\ \bibnamefont
  {Harris}}\ and\ \bibinfo {author} {\bibfnamefont {R.~V.}\ \bibnamefont
  {Lange}},\ }\href {\doibase 10.1103/PhysRev.157.295} {\bibfield  {journal}
  {\bibinfo  {journal} {Phys. Rev.}\ }\textbf {\bibinfo {volume} {157}},\
  \bibinfo {pages} {295} (\bibinfo {year} {1967})}\BibitemShut {NoStop}%
\bibitem [{\citenamefont {van Dongen}(1994)}]{vanDongen94}%
  \BibitemOpen
  \bibfield  {author} {\bibinfo {author} {\bibfnamefont {P.~G.~J.}\
  \bibnamefont {van Dongen}},\ }\href {\doibase 10.1103/PhysRevB.49.7904}
  {\bibfield  {journal} {\bibinfo  {journal} {Phys. Rev. B}\ }\textbf {\bibinfo
  {volume} {49}},\ \bibinfo {pages} {7904} (\bibinfo {year}
  {1994})}\BibitemShut {NoStop}%
\bibitem [{\citenamefont {K\"uhner}\ \emph {et~al.}(2000)\citenamefont
  {K\"uhner}, \citenamefont {White},\ and\ \citenamefont {Monien}}]{KWM00}%
  \BibitemOpen
  \bibfield  {author} {\bibinfo {author} {\bibfnamefont {T.~D.}\ \bibnamefont
  {K\"uhner}}, \bibinfo {author} {\bibfnamefont {S.~R.}\ \bibnamefont {White}},
  \ and\ \bibinfo {author} {\bibfnamefont {H.}~\bibnamefont {Monien}},\ }\href
  {\doibase 10.1103/PhysRevB.61.12474} {\bibfield  {journal} {\bibinfo
  {journal} {Phys. Rev. B}\ }\textbf {\bibinfo {volume} {61}},\ \bibinfo
  {pages} {12474} (\bibinfo {year} {2000})}\BibitemShut {NoStop}%
\bibitem [{\citenamefont {Ejima}\ \emph {et~al.}(2011)\citenamefont {Ejima},
  \citenamefont {Fehske},\ and\ \citenamefont {Gebhard}}]{EFG11}%
  \BibitemOpen
  \bibfield  {author} {\bibinfo {author} {\bibfnamefont {S.}~\bibnamefont
  {Ejima}}, \bibinfo {author} {\bibfnamefont {H.}~\bibnamefont {Fehske}}, \
  and\ \bibinfo {author} {\bibfnamefont {F.}~\bibnamefont {Gebhard}},\ }\href
  {http://stacks.iop.org/0295-5075/93/i=3/a=30002} {\bibfield  {journal}
  {\bibinfo  {journal} {EPL (Europhysics Letters)}\ }\textbf {\bibinfo {volume}
  {93}},\ \bibinfo {pages} {30002} (\bibinfo {year} {2011})}\BibitemShut
  {NoStop}%
\bibitem [{\citenamefont {Hao}\ \emph {et~al.}(2008)\citenamefont {Hao},
  \citenamefont {Zhang},\ and\ \citenamefont {Chen}}]{HZC08}%
  \BibitemOpen
  \bibfield  {author} {\bibinfo {author} {\bibfnamefont {Y.}~\bibnamefont
  {Hao}}, \bibinfo {author} {\bibfnamefont {Y.}~\bibnamefont {Zhang}}, \ and\
  \bibinfo {author} {\bibfnamefont {S.}~\bibnamefont {Chen}},\ }\href {\doibase
  10.1103/PhysRevA.78.023631} {\bibfield  {journal} {\bibinfo  {journal} {Phys.
  Rev. A}\ }\textbf {\bibinfo {volume} {78}},\ \bibinfo {pages} {023631}
  (\bibinfo {year} {2008})}\BibitemShut {NoStop}%
\bibitem [{\citenamefont {Freericks}\ \emph {et~al.}(2009)\citenamefont
  {Freericks}, \citenamefont {Krishnamurthy}, \citenamefont {Kato},
  \citenamefont {Kawashima},\ and\ \citenamefont {Trivedi}}]{FKKKT09}%
  \BibitemOpen
  \bibfield  {author} {\bibinfo {author} {\bibfnamefont {J.~K.}\ \bibnamefont
  {Freericks}}, \bibinfo {author} {\bibfnamefont {H.~R.}\ \bibnamefont
  {Krishnamurthy}}, \bibinfo {author} {\bibfnamefont {Y.}~\bibnamefont {Kato}},
  \bibinfo {author} {\bibfnamefont {N.}~\bibnamefont {Kawashima}}, \ and\
  \bibinfo {author} {\bibfnamefont {N.}~\bibnamefont {Trivedi}},\ }\href
  {\doibase 10.1103/PhysRevA.79.053631} {\bibfield  {journal} {\bibinfo
  {journal} {Phys. Rev. A}\ }\textbf {\bibinfo {volume} {79}},\ \bibinfo
  {pages} {053631} (\bibinfo {year} {2009})}\BibitemShut {NoStop}%
\bibitem [{\citenamefont {Lange}\ \emph {et~al.}(2017)\citenamefont {Lange},
  \citenamefont {Ejima},\ and\ \citenamefont {Fehske}}]{LEF17}%
  \BibitemOpen
  \bibfield  {author} {\bibinfo {author} {\bibfnamefont {F.}~\bibnamefont
  {Lange}}, \bibinfo {author} {\bibfnamefont {S.}~\bibnamefont {Ejima}}, \ and\
  \bibinfo {author} {\bibfnamefont {H.}~\bibnamefont {Fehske}},\ }\href
  {\doibase 10.1103/PhysRevLett.118.120401} {\bibfield  {journal} {\bibinfo
  {journal} {Phys. Rev. Lett.}\ }\textbf {\bibinfo {volume} {118}},\ \bibinfo
  {pages} {120401} (\bibinfo {year} {2017})}\BibitemShut {NoStop}%
\bibitem [{\citenamefont {Ejima}\ \emph
  {et~al.}(2012{\natexlab{b}})\citenamefont {Ejima}, \citenamefont {Fehske},\
  and\ \citenamefont {Gebhard}}]{EFG12}%
  \BibitemOpen
  \bibfield  {author} {\bibinfo {author} {\bibfnamefont {S.}~\bibnamefont
  {Ejima}}, \bibinfo {author} {\bibfnamefont {H.}~\bibnamefont {Fehske}}, \
  and\ \bibinfo {author} {\bibfnamefont {F.}~\bibnamefont {Gebhard}},\ }\href
  {http://stacks.iop.org/1742-6596/391/i=1/a=012143} {\bibfield  {journal}
  {\bibinfo  {journal} {Journal of Physics: Conference Series}\ }\textbf
  {\bibinfo {volume} {391}},\ \bibinfo {pages} {012143} (\bibinfo {year}
  {2012}{\natexlab{b}})}\BibitemShut {NoStop}%
\bibitem [{\citenamefont {zu~M\"unster}\ \emph {et~al.}(2014)\citenamefont
  {zu~M\"unster}, \citenamefont {Gebhard}, \citenamefont {Ejima},\ and\
  \citenamefont {Fehske}}]{zMGEF14}%
  \BibitemOpen
  \bibfield  {author} {\bibinfo {author} {\bibfnamefont {K.}~\bibnamefont
  {zu~M\"unster}}, \bibinfo {author} {\bibfnamefont {F.}~\bibnamefont
  {Gebhard}}, \bibinfo {author} {\bibfnamefont {S.}~\bibnamefont {Ejima}}, \
  and\ \bibinfo {author} {\bibfnamefont {H.}~\bibnamefont {Fehske}},\ }\href
  {\doibase 10.1103/PhysRevA.89.063623} {\bibfield  {journal} {\bibinfo
  {journal} {Phys. Rev. A}\ }\textbf {\bibinfo {volume} {89}},\ \bibinfo
  {pages} {063623} (\bibinfo {year} {2014})}\BibitemShut {NoStop}%
\bibitem [{ITe()}]{ITensor}%
  \BibitemOpen
  \href@noop {} {}\bibinfo {note} {{h}ttp://itensor.org/}\BibitemShut {NoStop}%
\end{thebibliography}%
\bibliographystyle{apsrev4-1}

\end{document}